%% file: main.tex
\documentclass[runningheads]{llncs}

\usepackage{geometry}
\usepackage[pdftex]{graphicx}

\usepackage{color}
\usepackage{xcolor,colortbl}
\usepackage{subfigure}
\usepackage{booktabs}
\usepackage{tabularx}
\usepackage{rotating}
\usepackage{multirow}
\usepackage{enumitem}
\usepackage{appendix}
\usepackage{enumerate}
\usepackage{amssymb}
\usepackage{amsmath}
\usepackage{amssymb,stmaryrd}
\usepackage{url}

\definecolor{fgreen}{rgb}{0.07, 0.53, 0.03}
\definecolor{grey}{gray}{0.6}
\definecolor{lgray}{gray}{0.85}
\definecolor{dgray}{gray}{0.6}

\input{macros}

\begin{document}


\title{Probabilistic Formal Modelling to Uncover and Interpret Interaction Styles}

\author{Oana Andrei, Muffy Calder, Matthew Chalmers, and Alistair Morrison}


\authorrunning{Andrei, Calder, Chalmers, Morrison}

\institute{School of Computing Science, University of Glasgow, UK \\ \email{firstname.lastname@glasgow.ac.uk}}

\maketitle

\begin{abstract}
We present a study using new computational methods, based on a novel combination of machine learning for inferring admixture hidden Markov models and probabilistic model checking, to uncover interaction styles in a mobile app.  These styles  are then  used to inform a redesign,  which is implemented, deployed, and then analysed using the same methods. The data sets are logged user traces, collected over two six-month deployments of each version, involving thousands of users and segmented into different time intervals. The methods  do not assume tasks or absolute metrics such as measures of engagement, but uncover the styles through unsupervised inference of clusters and analysis with probabilistic temporal logic.   For both versions there was a clear distinction between the styles adopted by users during the first day/week/month of usage, and during the second and third months, a result we had not anticipated.       
\end{abstract}

\section{Introduction}
\input{intro}

\section{Study Design}\label{sect:study}

\input{study}

\section{Study Results}\label{sect:results}
\input{results}

\section{Discussion}\label{sect:discussion}

\input{discussion}

\section{Related work}\label{sect:related}
\input{related}

\section{Conclusions}\label{sect:conclusions}
\input{conclusions}

\vspace{.3cm}
\noindent
{\bf Acknowledgements.} 
This research was supported by UKRI-EPSRC programme grants  EP/J007617/1 {\em A Population Approach to Ubicomp System Design} and  EP/N007565 {\em Science of Sensor System Software}.
 
\bibliographystyle{abbrv}
\bibliography{tochiBib}

\appendix

\input{appendix_AppTracker}

\input{pctl_background}

\input{appendix}

\end{document}

%% file: macros.tex
\newcommand{\opstate}{\mathsf{state}}

\newcommand{\propertyName}[1]{\texttt{#1}}

\newcommand{\VisitProbInit}{\propertyName{VisitProbInit}}
\newcommand{\StepCountInit}{\propertyName{StepCountInit}}
\newcommand{\VisitCountInit}{\propertyName{VisitCountInit}}
\newcommand{\VisitProbBetween}{\propertyName{VisitProbBtw}}
\newcommand{\StepCountBetween}{\propertyName{StepCountBtw}}
\newcommand{\SessionCount}{\propertyName{SessionCount}}
\newcommand{\SessionLength}{\propertyName{SessionLength}}
\newcommand{\StateToEndProb}{\propertyName{StateToStop}}
\newcommand{\StateToAPProb}{\propertyName{StateToPattern}}
\newcommand{\StateToPattern}{\propertyName{StateToPattern}}
\newcommand{\LongRunPattern}{\propertyName{LongRunPattern}}

\newcommand{\patternlabel}[1]{\texttt{#1}}
\newcommand{\Glancing}{\patternlabel{Glancing}}

\newcommand{\Browsing}{\patternlabel{Browsing}}
\newcommand{\Focussing}{\patternlabel{Focussing}}
\newcommand{\Summary}{\patternlabel{Summary}}
\newcommand{\Specific}{\patternlabel{Specific}}
\newcommand{\Session}{\patternlabel{Session}}

\newcommand{\MyTopApps}{\patternlabel{Summary}}
\newcommand{\SpecificByPeriod}{\patternlabel{SpecificByPeriod}}
\newcommand{\SpecificByApp}{\patternlabel{SpecificByApp}}
\newcommand{\SpecificStats}{\patternlabel{SpecificStats}}
\newcommand{\SpecificMenu}{\patternlabel{SpecificMenu}}
\newcommand{\TopLevelMenu}{\patternlabel{TopLevelMenu}}

\newcommand{\LStates}{\mathcal{X}}
\newcommand{\OStates}{\mathcal{Y}}

\ifx\property\undefined
\newtheorem{property}{Property}
\fi

\newcommand{\true}{\mathtt{true}}

\newcommand{\Probs}{\mathtt{P}}
\newcommand{\Probss}{\mathtt{S}}
\newcommand{\Probr}{\mathtt{R}}

\newcommand{\D}{\mathcal{D}}

\newcommand{\filter}{\mathit{filter}}

\newcommand{\TG}{{\sf G}}
\newcommand{\TF}{{\sf F}}
\newcommand{\TU}{{\sf U}}

\newcommand{\TX}{{\sf X}}
\newcommand{\TC}{{\sf C}}

\newcommand{\statelabel}[1]{\texttt{#1}}

\newcommand{\TaC}{\statelabel{T\&C}}
\newcommand{\UseStart}{\statelabel{UseStart}}
\newcommand{\Main}{\statelabel{Main}}
\newcommand{\OverallUsage}{\statelabel{OverallUsage}}

\newcommand{\LastSevenDays}{\statelabel{Last7Days}}
\newcommand{\PeriodSelector}{\statelabel{SelectPeriod}}
\newcommand{\SelectPeriod}{\statelabel{SelectPeriod}}
\newcommand{\AppsInPeriod}{\statelabel{AppsInPeriod}}
\newcommand{\Settings}{\statelabel{Settings}}
\newcommand{\UseStop}{\statelabel{UseStop}}
\newcommand{\Stats}{\statelabel{Stats}}

\newcommand{\UBCOverallUsage}{\statelabel{ChartOverall}}

\newcommand{\UBCStats}{\statelabel{ChartStats}}

\newcommand{\Feedback}{\statelabel{Feedback}}
\newcommand{\UsageBarChartAppsInPeriod}{\statelabel{ChartAppsInPeriod}}

\newcommand{\Info}{\statelabel{Info}}
\newcommand{\Task}{\statelabel{Task}}

\newcommand{\OverallAll}{\statelabel{OverallAll}}
\newcommand{\ExploreData}{\statelabel{ExploreData}}
\newcommand{\AppsToday}{\statelabel{AppsToday}}
\newcommand{\AppsbyPeriod}{\statelabel{AppsbyPeriod}}
\newcommand{\UBCAppsInPeriod}{\statelabel{ChartAppsInPeriod}}
\newcommand{\ChartAppsInPeriod}{\statelabel{ChartAppsInPeriod}}
\newcommand{\OverallbyApp}{\statelabel{OverallbyApp}}
\newcommand{\UBCOverallbyApp}{\statelabel{ChartOverallbyApp}}
\newcommand{\ChartOverallbyApp}{\statelabel{ChartOverallbyApp}}

\newcommand{\startLabel}{\statelabel{startS}}
\newcommand{\stopLabel}{\statelabel{stopS}}



%% file: intro.tex
Menu driven interfaces are often designed to fulfill perceived or
expected end user needs and interaction styles, yet in practice users
may adopt numerous styles, some of them unanticipated by designers.
This may be due to many factors, including software
appropriation~\cite{Dix07,Dourish,Tchounikine}, the time since they
started using the software (e.g.~the first week or after six months
usage), intentions concerning that particular use (e.g.~a quick use,
or a long and thorough use), or state of mind (e.g.~distracted or
intentional), etc.  The result is interaction styles can vary both
from user to user~\cite{ZhaoR16} and, over time, for each individual
user, within and between interaction sessions.  Consequently,
designers may wish to redesign an interface in the light of how users
have been using their system over time.

Current tools for studying interaction include qualitative methods,
such as interviews, think-alouds, and direct observations, can help
uncover users' behaviours and preferences, but these procedures are
expensive to carry out with large user populations.  Restricting to
smaller sample sizes might miss some forms of activity, and might also
be biased culturally, e.g., if based on users local to the designers.
Quantitative methods can include many more users --- maybe literally
{\em all} the users --- but most existing methods rely on modelling
assumptions made in advance, e.g. assumed tasks that users carry out,
or on relatively simple metrics. For example, they may focus on the
statistics of occurrence of basic features such as time in app and
screens visited, or on task-based measures~\cite{ferre2017extending}.

We describe a   study of user-centred redesign of a mobile
app, that uses a new quantitative approach for studying
interaction. The data are logged user interaction traces, extracted
from a population of app users (in our case, all users).
We segmented the traces into several data sets according the time
intervals: 1st day, 1st week, 1st, 2nd, and 3rd month of usage, so
we could observe if styles correlate with length of engagement. We did
not pre-suppose tasks, but logged all interactions that change user
views. From these data sets we inferred computational models using
machine learning (ML) unsupervised clustering methods.  A novelty of
our approach is that we used probabilistic temporal logic properties
to interrogate the inferred models, and then interpreted the results,
using inductive coding, to evaluate how well the design supports the
interaction styles we uncovered.  This informed a redesign of the
interface, which we then deployed and studied in the same way.

The study involved two design iterations of the hierarchical menu interface
for AppTracker~\cite{MorrisonXHBC18}, which allows its users to keep
track of the usage of their device, and can be thought of as an
instrument for \textit{personal informatics}~\cite{Wolf09,LiDF10}.
Each design, which we call AppTracker 1 and AppTracker2, was deployed
for at least 6 months, involving thousands of users. The work
presented here represents a rare and long term collaboration over
several years, between researchers in human-computer interaction
evaluation, mobile app design, machine learning, and model checking.

The paper is organised as follows.  The next section contains an
overview of AppTracker and the study design.  In
Sect.~\ref{sect:results} we give the main results, which include the
interaction styles uncovered in AppTracker1, the redesigned interface
for AppTracker2, the interaction styles uncovered in AppTracker2, and a
comparison of interaction styles uncovered in AppTracker1 and
AppTracker2.  We reflect on our findings and the role of ML in
Sect.~\ref{sect:discussion}, related   work is in
Sect.~\ref{sect:related}, and conclusions and future work in
Sect.~\ref{sect:conclusions}.

%% file: study.tex
AppTracker  allows users to keep track of and view the
usage of all mobile apps on their iPhone or iPad device. It
runs in the background, monitoring the opening and closing of all apps
installed on the device, as well as every time the device is locked or
unlocked.  It collects this data and can then display a series of
charts and statistics, offering users insight into their behaviour,
such as time spent per day on their device, or their most used apps
over time.  The user interface is based on hierarchical
menus, allowing navigation through the menu structure to access charts
and summary statistics. AppTracker generates two distinct forms of
data to study: i) a user's use of their device, with records of every
app they launch, and ii) user interactions {\em within the AppTracker
  application itself} such as button clicks and screen changes.  The
former type of data has been analysed in previous
work~\cite{MorrisonXHBC18}, this study is based on the latter type of data.

\subsection{Overview}
\begin{figure*}[!t]
\centering
\includegraphics[width=0.7\textwidth]{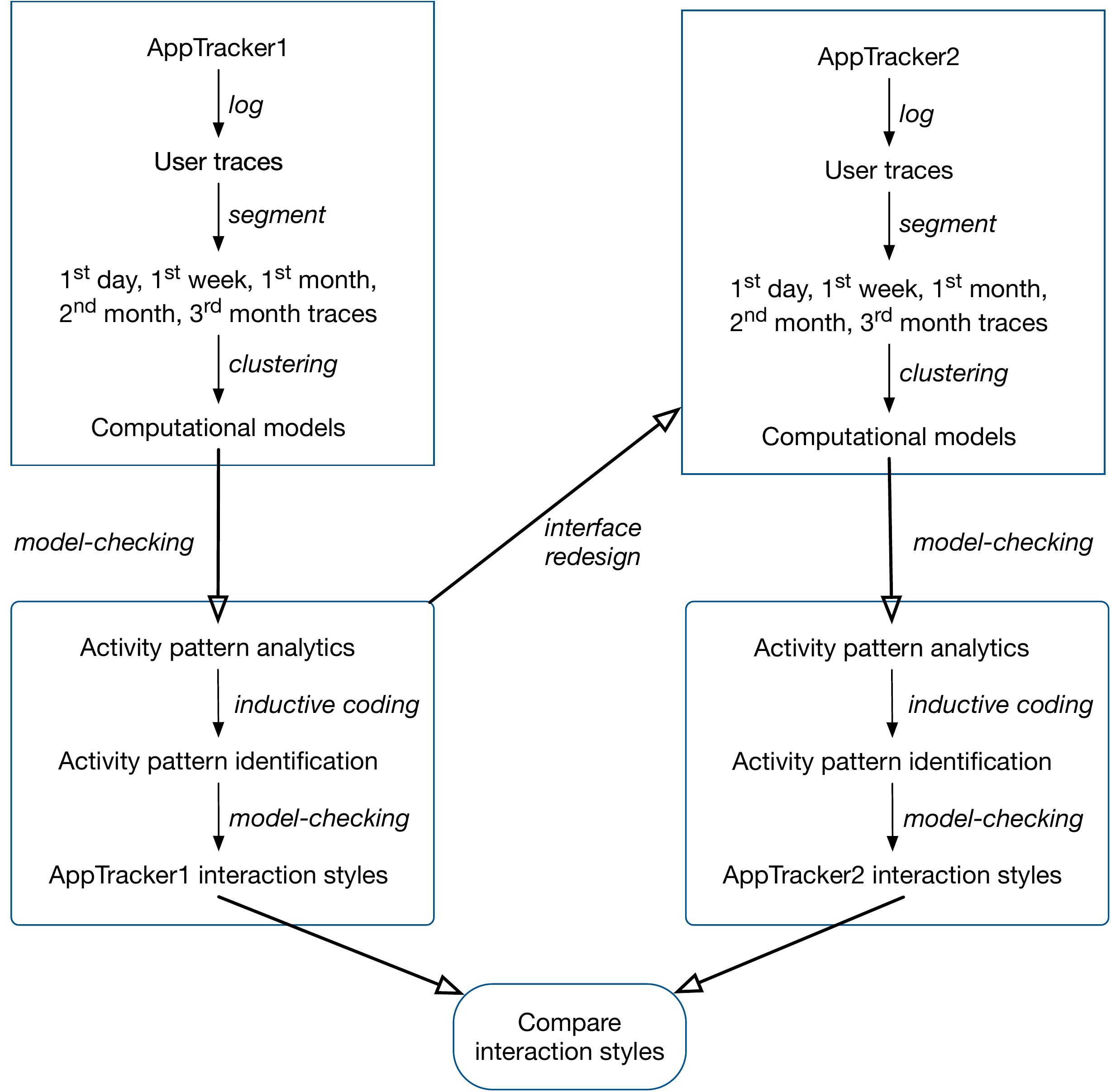}
  \caption{Study design}
\label{fig:Study_Design}
\end{figure*}

Figure~\ref{fig:Study_Design} outlines the flow of the study.
AppTracker1 user interactions were logged  across a large  population of users, 
traces were segmented into time series data sets,   and admixture Markov models  were inferred from those data sets using an ML clustering algorithm. 
A key concept for each Markov model is the inferred {\em activity patterns} and probabilities to transition between them; together these   
encapsulate common observed temporal behaviours  shared across a set of
logged user traces. Each activity pattern is a discrete-time Markov
chain, in which observed variables label the AppTracker states and each  
pattern corresponds to a latent state in the admixture Markov model of
all interaction behaviour.
We analysed the models' activity patterns using a set of (parameterised)  probabilistic temporal logic properties and model checking~\cite{BaierKatoen-MCbook}. We used inductive coding~\cite{Thomas06} to categorise the results, e.g.~numbers of steps to reach a state, session lengths, predominant states, and then  interpreted the results to identify the activity patterns. We used another set of temporal logic properties to produce  the long run likelihoods and probabilities to transition between activity patterns,  all of which contributes to a description of the interaction styles. 
We considered how well AppTracker1 supported users interaction styles and addressed deficiencies in a redesign to feed into the next version AppTracker2 which was then deployed. We iterated the analytics again, based on AppTracker2 user traces, and concluded the study by comparing the interaction styles in AppTracker1 and AppTracker2.  

\subsection{Time Series Data}

 Each logged user trace   is a sequence of event labels.   Each trace consists of many user interaction sessions,    which start  when the  
application is launched or being brought to the foreground (denoted by event  \startLabel)
 and
end  when the application is closed or put in the background (denoted by event  \stopLabel).
 User traces are segmented by  time intervals $[t_1,t_2]$ such that
the first session of the segment occurs on or after time-stamp $t_1$
and the last session of the fragment occurs before time-stamp $t_2$; the whole trace may extend beyond time-stamp $t_2$.

\subsection{Interface for AppTracker1 }

\begin{figure*}[!t]
  \centering 
  \subfigure[Main menu]{\includegraphics[scale=0.128]{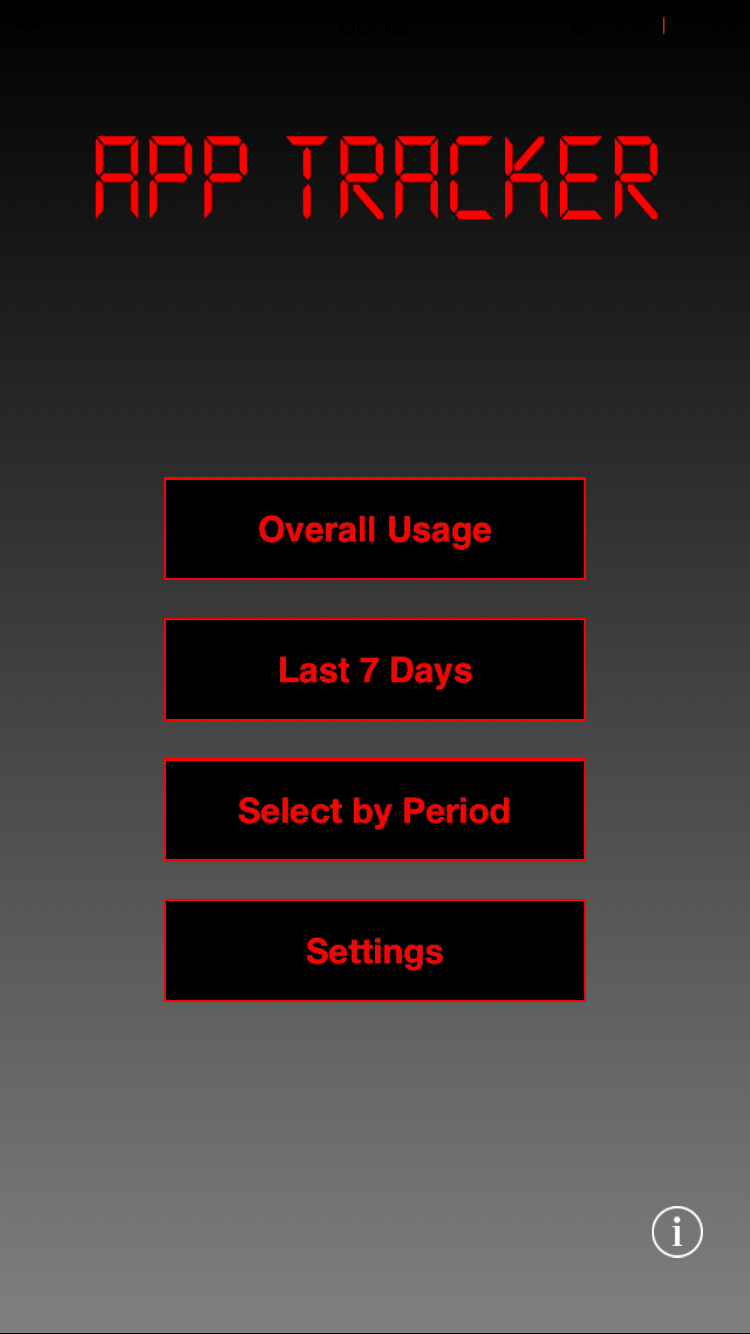}\label{fig:menu}}
  \quad
  \subfigure[Stats]{\includegraphics[scale=0.15]{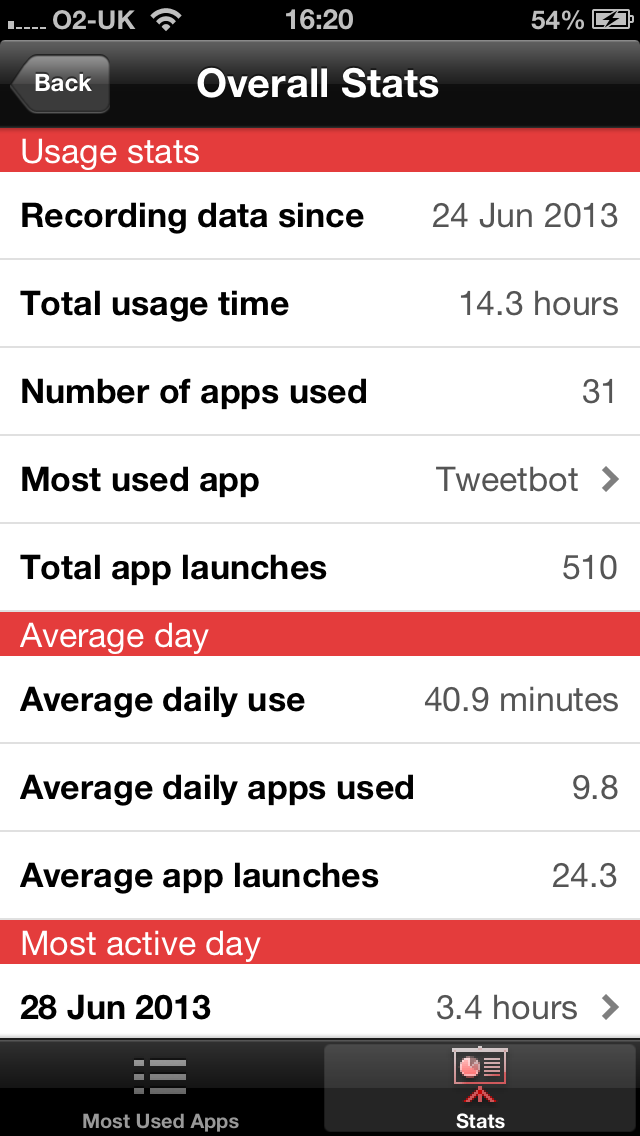}\label{fig:stats}}
  \quad  
  \subfigure[Daily device usage]{\includegraphics[scale=0.15]{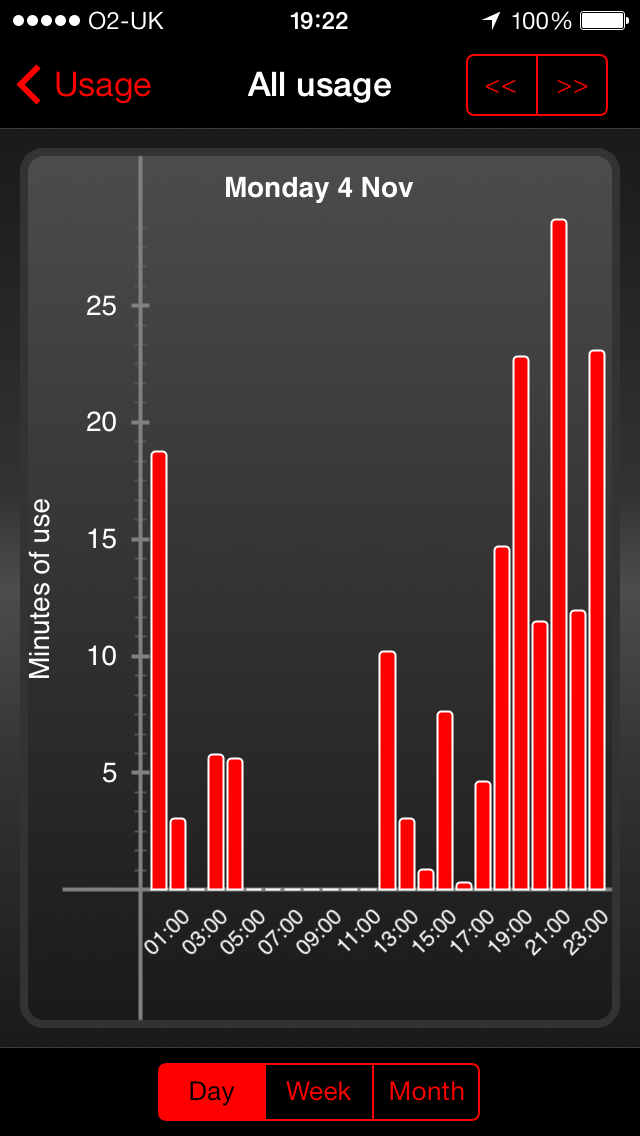}\label{fig:day}} 
  \caption{Screenshots from AppTracker1: (a) The main menu view
    corresponds to the state \Main; (b) The stats view corresponds to
    the state \Stats. (c) The daily device usage view corresponds to
    the state \UBCOverallUsage\ when the selected period is Monday 4
    November.}
\label{fig:screenshotsAT1}
\end{figure*}

\begin{figure*}[!ht]
 \centering 
 \includegraphics[scale=0.55]{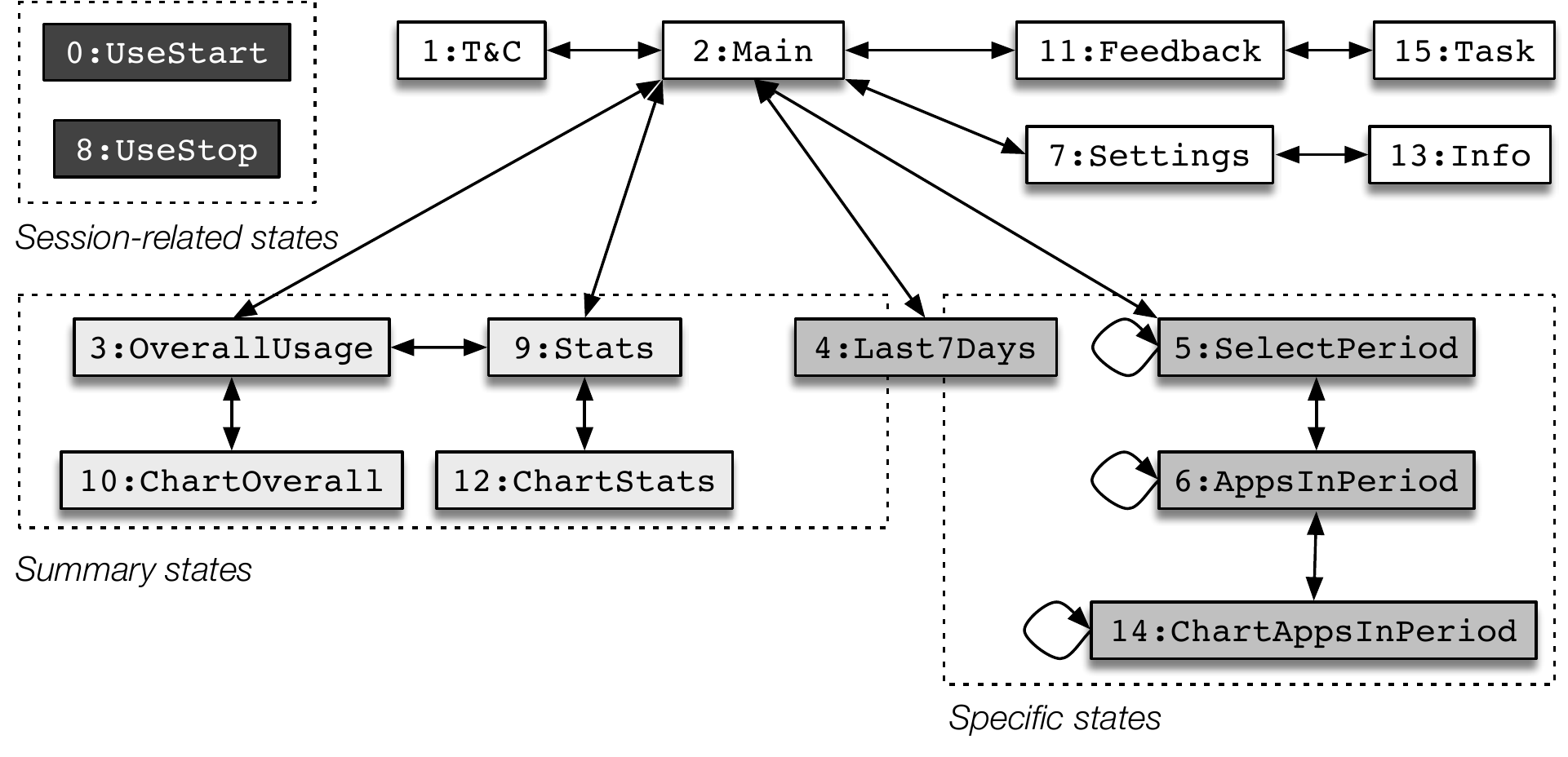}
 
\caption{ AppTracker1   state diagram. The vertical
    vertical layout of the menus corresponds to a
    left to right traversal of the states in the dotted boxes.}
    \label{fig:stateDiagramAT1}
\end{figure*}

On launching AppTracker1, the main menu screen  
offers four main options (Fig.~\ref{fig:menu}); from top to bottom
they are:
\begin{itemize}
\item {\em Overall Usage} contains summaries of all the data recorded
  since AppTracker1 was installed and opens the views \OverallUsage\
  and \Stats\ (Fig.~\ref{fig:stats}).  
\item {\em Last 7 Days} opens the view \LastSevenDays\ and displays a
  chart of activity of the user's 5 most used apps during the last 7 days. 
\item {\em Select by Period} opens the view \PeriodSelector\ and shows
  statistics for a selected period of time, e.g.~which apps were used
  most since Friday, the daily time spent on Facebook over the last
  month, hourly device usage on Monday (Fig.~\ref{fig:day}).
\item {\em Settings} allows a user to start and stop the tracker, or
  to reset their recorded data.
\end{itemize}
There are 16 user-initiated events that switch between views (the name of the event
is the resulting view), see state diagram in Fig.~\ref{fig:stateDiagramAT1}.
States are grouped into \Summary\ states for viewing summary or overall usage data; 
\Specific\ states for viewing drilled down, specific data; \Session-related\ states
marking the start and the end of a session; and remaining states. Note state 
\LastSevenDays\ is a summary state or a specific state depending on the context of use.

Logged interaction data are stored in a MySQL database using the SGLog
framework~\cite{HallBMRSC09} and processed using JavaScript to obtain user 
traces in JSON format. AppTracker1 was first released in August 2013 and downloaded over
35,000 times.
Our data sets are taken from a sample of 322 user traces
during 2013 and 2014. The maximum session count over all the traces is
129, the minimum was limited to 5.

\subsection{Computational Methods}

We used ML clustering methods to infer admixture Markov models, first defined in~\cite{AndreiC18}, based on first-order auto-regressive hidden Markov 
Models (AR-HMM)~\cite{Murphy12}. Admixture models permit interleaved variation, 
so we can model users that may switch interaction styles both {\em within}  and {\em between}  
sessions. We include some definitions here for completeness. 

A {\em discrete-time Markov chain} (DTMC) is a tuple
$\D=(S,s_0,\Probs,\mathcal{L})$ where: $S$ is a set of states; $s_0\in
S$ is the initial state; $\Probs:S\times S\rightarrow [0,1]$ is the
transition probability function such that for all states
$s\in S$ we have $\sum_{s'\in S}\Probs(s,s')=1$; and $\mathcal{L}:S\rightarrow
2^\mathcal{A}$ is a labelling function associating to each state $s$
in $S$ a set of valid atomic propositions from a set $\mathcal{A}$.  A
{\em path} (or execution) of a DTMC is a non-empty sequence
$s_0s_1s_2\ldots$ where $s_i\in S$ and $\Probs(s_i,s_{i+1})>0$ for all
$i\geq 0$.  A transition is also called a {\em time-step}.

A {\em first-order auto-regressive hidden Markov 
Model} (AR-HMM)~\cite{Murphy12} is as a tuple
$(\LStates,\OStates,\pi,A,B)$ where: $\LStates$ is the set of hidden
(or latent) states \mbox{$\LStates=\{1,\ldots,K\}$;} $\OStates$ is the set of
observed states generated by hidden states; $\pi:\LStates \rightarrow
[0,1]$ is an initial distribution with $\sum_{x \in
  \LStates}\pi(x)=1$; $A:\LStates \times \LStates \rightarrow [0,1]$
is the transition probability matrix, such that for all $x \in
\LStates$ we have $\sum_{x' \in \LStates}A(x,x')=1$; $B : \LStates
\times \OStates \times \OStates \rightarrow [0,1]$ is the observation
probability matrix, such that for all $x \in \LStates$ and $y \in
\OStates$ we have $\sum_{y' \in \OStates}B(x,y,y')=1$.

\begin{figure*}[!t]
  \centering 
  \includegraphics[width=0.55\textwidth]{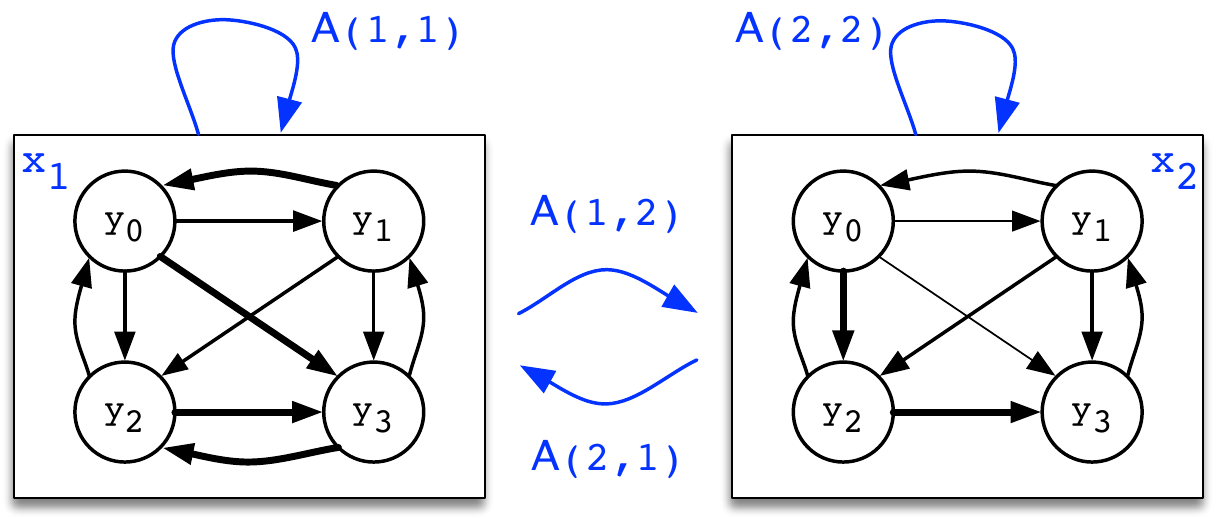}
  \caption{Pictorial representation of a GPAM(2) with the
    latent states $x_1$ and $x_2$. The two activity patterns are the
    DTMCs in each box; there are four observed states $y_0,y_1,y_2,y_3$.
    Transition probabilities are indicated by the thickness of
    transitions.}
\label{fig:admixture_model_gpam2}
\end{figure*}

Now let $\mathcal{P}$ be a
population of $M$ user traces over $n$ different types of event
labels, $\mathcal{A}$ the set of the labels of all events occurring in
$\mathcal{P}$, and $K$ a positive integer.  A {\bf generalised
  population admixture model with $K$ components} or {\bf GPAM($K$)} for the
user trace population $\mathcal{P}$ is a tuple $(\LStates, \OStates,
\pi, A, B, \mathcal{L})$ where $(\LStates,\OStates,\pi,A,B)$ is an
AR-HMM with $|\LStates|=K$ and $|\OStates|=n$, and
$\mathcal{L}:\OStates \rightarrow {\mathcal A}$ is the labelling
function mapping a unique event name to each observed state in
$\OStates$. A pictorial representation of a GPAM(2) is given in
Fig.~\ref{fig:admixture_model_gpam2}.  For any GPAM $(\LStates,
\OStates, \pi, A, B, \mathcal{L})$ and a latent state $x\in \LStates$,
the tuple $(\OStates,\mathcal{L}^{-1}(\startLabel), B(x),\mathcal{L})$
is a discrete-time Markov chain (DTMC) called an {\bf activity
  pattern}. 
  
For each input data set $\mathcal{P}$ of user traces, we compute
$n\times n$ transition-occurrence matrices for each trace $\alpha$
such that matrix position $(i,j)$ is the number of times the
subsequence $\alpha_{i} \alpha_{j}$ (i.e.~two adjacent event labels)
occurs in $\alpha$. The resulting transition-occurrence matrices are
the input data for ML; we employ the Baum—Welch clustering
algorithm~\cite{BaumWelch}, which uses the local non-linear
optimisation Expectation--Maximisation (EM) algorithm~\cite{Demp1977}
for finding maximum likelihood parameters of observing each trace, and
restarting the algorithm whenever the log-likelihood has
multiple-local maxima. 

We used the probabilistic model checker PRISM~\cite{KwiatkowskaNP11} to analyse temporal
properties expressed in rPCTL, an extension of Probabilistic Computation Tree Logic  PCTL*~\cite{KwiatkowskaNP11,BaierKatoen-MCbook} with rewards; an overview  
can be found in Appendix~\ref{pctl_background}. 

The set of parameterised properties we defined and used is contained in Table~\ref{tbl:properties}. These temporal properties express behaviour {\em within} an activity pattern or behaviour that involves {\em several} activity patterns. The properties of the latter type are \StateToAPProb\ and \LongRunPattern, and we used them for model checking on GPAMs only. The rest of the properties were used on activity patterns as DTMCs. 
First we model-check the temporal properties \VisitProbInit, \StepCountInit,
\VisitCountInit, \SessionLength, and \SessionCount\ for different, incremental values of $K$ for all states.  Optionally, we use the properties \VisitProbBetween\ and \StepCountBetween\ when it is too difficult to interpret the results from the other properties on activity patterns.
Then the properties \StateToAPProb\ and \LongRunPattern\ help us identify more nuanced characteristics by considering if an activity pattern changes within a session and the long run probability of each activity pattern, respectively. Other PTCL* properties considered, but not included in this paper, helped us analyse whether some particular states lead to the end of a session in fewer steps than other states and whether there are correlations between state formulae (over states and/or patterns).

\begin{table}[t!]
\caption{Reward-based PCTL (rPCTL) properties are parameterised by: $N$ a positive integer value for the number of time steps; $j,j_1, j_2$ positive integer values denoting observed state identifiers; $i,i_1,i_2$  integer values between 1 and $K$ denoting the hidden state identifiers (activity patterns); $p\in [0,1]$ a probability. }
\label{tbl:properties}
{\small
    \begin{tabularx}{\textwidth}{lX}
      \toprule
      \textbf{Name} & \textbf{Formula and informal description} \\
      \midrule \VisitProbInit & $\Probs_{=?}[\true\, \TU^{\leq
        N}(y=j)]$ : probability to reach $j$ from the initial state
      within $N$ steps
      \\
      \StepCountInit & $\Probr_{\texttt{rSteps}=?}[\TF\, (y=j)]$ :
      expected number of steps to reach $j$ from the initial state\\
      \VisitCountInit & $\Probr_{\texttt{rStatej}=?}[\TC^{\leq N}]$ :
      expected number of visits to $j$ within $N$ steps
      \\
      \midrule 
      \SessionLength & $\Probr_{\texttt{rSteps}=?}[\TF\, (y=
      \stopLabel)]$ : expected number of steps until the end of session.\\
      \SessionCount & $\Probr_{\texttt{rState\stopLabel}=?}[\TC^{\leq
        N}]$ : expected number of sessions within $N$ steps 
      \\
      \midrule \VisitProbBetween &
      $\filter(\opstate,\Probs_{=?}[(\lnot\stopLabel)\, \TU^{\leq
        N}(y=j_2)],(y=j_1))$ : probability of reaching observed state
      $j_2$ from $j_1$ within the same session 
      \\
      \StepCountBetween & $\filter(\opstate,
      \Probr_{\texttt{rSteps}=?}[\TF\, (y=j_2)], (y=j_1))$ : expected
      number of steps to reach state $j_2$ from $j_1$   \\
      \midrule 
      \StateToAPProb & $\Probs_{\geq 1}[\,\TF\,(x=i_1\land y=j)] \land
      \Probs_{\geq 1}[\,\TG\,((x=i_1\land y=j) \Rightarrow
      \Probs_{>p}[(x=i_1 \land \lnot \stopLabel) \,\TU\, (x= i_2)])]$:
      likelihood of observed state $j$ in activity pattern $i_1$
      leading to changing the
      activity pattern to $i_2$ within the same session   \\
      \StateToEndProb & $\Probs_{\geq 1}[\,\TF\,(x=i\land
      y=j)] \land \Probs_{\geq 1}[\,\TG\,((x=i\land y=j) \Rightarrow
      \Probs_{>p}[(x=i) \,\TU\, \stopLabel\,])]$ : likelihood of
      observed state $j$ in activity pattern $i$ leading to the end of
      a session, without changing the activity pattern 
      \\
      \LongRunPattern & $\Probss_{=?}[\, x=i\, ]$ :
      probability of being in activity pattern $i$ in the long run 
      \\
      \bottomrule
    \end{tabularx}
}
\end{table}

Observed states are grouped, according to their expected function and
purpose within the application, for example \Summary\ and \Specific\ for AppTracker1. This grouping is done iteratively: an initial grouping is proposed by the designers, then after analysing the results 
of the temporal properties within each activity pattern, the analysts and designers may, together,
revise the grouping. Note that such state groups may overlap. 

The source code for the quantitative analysis is available online~\footnote{\url{https://github.com/oanaandrei/temporalanalytics}}.

\subsection{Inductive Coding}

While the PRISM model checking results are quantitative, interpretation of those results is subjective, and therefore similar to the interpretation of qualitative data.
We adopt the general inductive approach for analysing qualitative
evaluation data~\cite{Thomas06} that aims to allow research findings
to emerge from the data, rather than a deductive approach that tests a
hypothesis. 

Coding was carried out independently by two evaluators (authors Andrei and Calder), with several revisions and refinements, then checked for clarity (all four authors), and stakeholder checks (Morrison and Chalmers as designers).    
We note that many hours were spent choosing meaningful labels.  Since
the labels themselves have inherent meaning, the labels were changed
several times throughout this work, as we gained a deeper
understanding of all the nuances of the activity patterns. We suggest
this is both a strength and a weakness of our approach.

%% file: results.tex
 We give an overview of results, following the study flow in Fig.~\ref{fig:Study_Design}. For brevity we report results only for GPAM(2) models; details for GPAM(3) models are available in Appendix~\ref{appendix_gpam3}. 
We refer to results for models from data sets for 1st day, 1st week, and 1st month as results for 
{\em early days usage} and for the 2nd and 3rd month as {\em experienced usage}.

For inferring the models, Baum-Welch algorithm was implemented  in Java and ran on a
2.8GHz Intel Xeon (single thread, one core); it was restarted 200 times with 100 maximum
number of iterations for each restart. As example performance, for the data set consisting of the first month of usage, the algorithm took 3.2 min for $K = 2$, 4.1 min for $K = 3$, and 5.3 min for $K = 4$.

We started by model checking the properties \VisitProbInit, \StepCountInit,
\VisitCountInit, \SessionLength, and \SessionCount\ for $K=2$ for all states.  We selected a single value for $N$, typically somewhere
between $10$ and $150$; in our experience $50$ is a good starting
value. Note that the result from the model checker may not be a number (either
probability or positive real), due to state unreachability, a filter
satisfying no states, or the iterative method not converging within
100,000 iterations (limit chosen for the PRISM model checker).  In
this case the result is given as "---". To aid interpretation,
results are ordered "best" to "worst" as follows: greatest to least value for \VisitProbInit, \VisitProbBetween, and \VisitCountInit, while least to greatest value for \StepCountInit\ and \linebreak \StepCountBetween. 
This ordering reflects the following judgments: a higher probability
to visit a state, a higher number of state visits, and fewer steps to
reach a state, are all indicators of greater (user) interest in a
state or a pair of states. We encode this ordering visually using
the colour \textcolor{blue}{blue} for the "best" results and \textcolor{purple}{purple} for the "worst"
results.

\subsection{AppTracker1 Interaction Styles}

Tables~\ref{table:AT1_GPAM2_SAP_InitProps_4statesL1} and~\ref{table:AT1_GPAM2_SAP_VP1_VC1_SC1_UseStop} include activity pattern results for  
the temporal properties \VisitProbInit, \VisitCountInit, and
\StepCountInit\ analysed on GPAM(2) models inferred from different time intervals of usage. 

We say that a state is {\em predominant} within an activity pattern
if: (i) the probability to reach it computed using \VisitProbInit\ is
greater than 0.5, the visit count computed by \VisitCountInit\ is
greater than 1, and the number of time steps to reach it computed by
\StepCountInit\ is lower than the time bound $N=50$ chosen for the
temporal properties, and (ii) there is no other activity pattern in
the same model where the \VisitCountInit\ and \StepCountInit\ score at
least three times better.  An activity pattern is {\em
  centred} on a set of states (or state grouping they belong to) when those states are predominant within
the activity pattern. 
 Activity patterns are labelled according to two dimensions: {\em usage
  intensity} and {\em predominant states}.  Values in a usage
intensity type are based on abstractions of the frequency and length
of session, as well as on results returned by the properties
\VisitProbInit, \VisitCountInit, \StepCountInit, \VisitProbBetween,
and \StepCountBetween\ for the predominant states in relation to the
intensity usage type.  

 Table~\ref{table:GPAM2_initial_AP_categ_AT1} shows our workings
through the inductive coding process:
\begin{itemize}
    \item First we consider
session characteristics and use the Jenks natural breaks optimisation
method~\cite{Jenks67} to determine the best arrangement of session
count values into three categories, and similarly for the session
length. 
\item Second, we give an initial categorisation of the
activity patterns based on session characteristics and predominant
states. We can see immediately there is a correlation between
fewer/longer sessions, and more numerous/shorter sessions. We note
that more many/long sessions and few/short sessions did not occur. The
state identifiers upon which each activity pattern is centred are
listed in decreasing order of their results: the better the result for
a state in a pattern (the higher probability to reach, the higher
visit count, the fewer steps to reach) the higher the predominance of
that particular state compared to other states. We refer to a subset of states as \TopLevelMenu, these are states that are reached in one or two button presses in average (corresponding to time steps in the respective DTMC). 
\item Third, we conclude that the usage intensity type consists of three values that we call
\Browsing, \Glancing, and \Focussing.
\item Finally, we combine usage intensity and predominant state group type to assign four activity pattern labels. Note that two possible combinations are not present: \Glancing\Summary\ and
\Browsing\Specific.
\end{itemize}

We can see a clear split between the activity patterns for early days usage
and experienced usage.  

\begin{figure*}[!t]
  \centering 
 \subfigure[AppTracker1]{\includegraphics[width=0.9\textwidth]{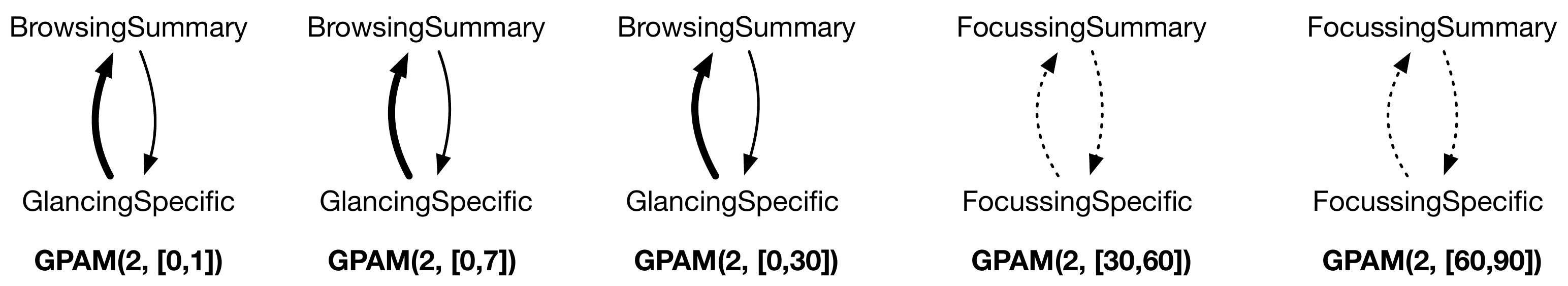}\label{fig:at1_gpam2_state2pattern}}
  \vspace{.9cm}
 \subfigure[AppTracker2]{\includegraphics[width=1\textwidth]{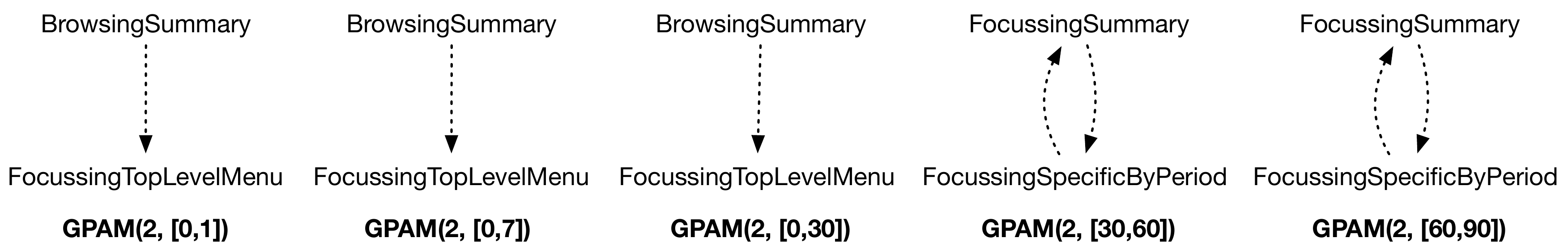}\label{fig:at2_gpam2_state2pattern}} 
 \caption{AppTracker1 and AppTracker2. Visualisations of
    the average likelihood of changing activity pattern
   from an   state within a session in GPAM(2) models for
   different time intervals. Excepting the \Stats\ and \UBCStats\
   states, for which the property returns lower probabilities than the
   average, the probabilities returned for all states are very
   similar. Dotted arrows denote probabilities in $[0.02,0.2)$, thin
   arrows in $[0.5,0.8)$, and thick arrows $\geq 0.8$. Likelihoods
   lower than 0.02 are omitted. These results are based on model checking the temporal property \StateToAPProb\ for various combinations of states and activity patterns.}
\label{fig:at1_at2_gpam2_state2pattern}
\end{figure*}

After labelling of the activity pattern, a further analysis based on model checking is carried out to investigate further the relationships between them using the temporal properties \LongRunPattern\ and \StateToAPProb, hence find out more about the interaction styles. 
The long run probabilities for activity patterns are shown in Fig.~\ref{fig:steadystate_AT1_AT2}(a) and probabilities that, for a given activity
pattern, a state leads to activity pattern change (within a session) 
are given in Fig.~\ref{fig:at1_at2_gpam2_state2pattern}(a).
The latter shows that for the early days usage, all states are highly likely (close to
probability 1) to transition from \Glancing\Specific\ to
\Browsing\Summary\ behaviour, 
and less likely the other way around (probability between 0.5 and
0.75), except for \Stats\ for which the probability is around 0.2, and
for \UBCOverallUsage\ and \UBCAppsInPeriod\ with even lower
probability. 
These two exceptions correlate with the results
highlighted in the \StateToEndProb\ analysis (not shown here) that indicate that once in
\Browsing\Summary\ in any of the states \Stats, \UBCOverallUsage, and
\UBCAppsInPeriod, it is unlikely to move to another activity pattern
before the end of session.  In the experienced user models, it is more
likely to move from \Focussing\Summary\ to \Focussing\Specific\ than
the other way around, but overall, it is unlikely to move between
these two patterns within a session.  

In conclusion, the interaction styles of AppTracker1 based on GPAM(2) models are summarised as follows. 
Early days usage contained \Glancing\ and \Browsing\ patterns, the latter possibly because users explore all the screens and features offered by the app.  For experienced users, usage is {\em focussed} with
\Focussing\Specific\ being a little more likely. \Glancing\ patterns involve the shortest sessions in terms of screen view counts over all patterns, and also appear as the most numerous sessions. We note similarity in these \Glancing\ patterns to the micro-usages discussed in~\cite{FerreiraGKBD14}, defined as "brief bursts of interaction with applications". 

 
\begin{table}[!t] 
  \center
  \caption{AppTracker1 activity pattern analytics. Results for the properties
    \VisitProbInit, \VisitCountInit, and \StepCountInit\ instantiated
    with $N=50$, the states \OverallUsage, \LastSevenDays,
    \SelectPeriod, \Stats, \AppsInPeriod, and the two activity
    patterns, which we call AP1 and AP2, on GPAM(2) models for five time
    intervals. Each row corresponds to a GPAM(2) model learned from the
    segmented data set for the respective time interval.  
    \textcolor{blue}{Blue} indicates best result and
    \textcolor{purple}{purple} indicates worst for a particular state
    across the activity patterns in a GPAM(2) model for a particular
    time interval.} 
    \label{table:AT1_GPAM2_SAP_InitProps_4statesL1} 

\begin{tabular}{|c|c|rr|rr|rr|rr|rr|}
  \hline
  \multirow{2}{*}{\begin{sideways}{\scriptsize\bf Prop.}\end{sideways}} & {\scriptsize\bf Time} & \multicolumn{2}{@{}c@{}|}{\bf
    \OverallUsage} & \multicolumn{2}{@{}c@{}|}{\bf
    \LastSevenDays} & \multicolumn{2}{@{}c@{}|}{\bf
    \SelectPeriod} & \multicolumn{2}{@{}c@{}|}{\bf
    \Stats} & \multicolumn{2}{@{}c@{}|}{\bf
    \AppsInPeriod}  \\   
  \cline{3-4} \cline{5-6} \cline{7-8}
  \cline{9-10} \cline{11-12}  & {\scriptsize\bf 
    interval} & {\scriptsize\bf AP1} & {\scriptsize\bf AP2} & {\scriptsize\bf AP1} & {\scriptsize\bf AP2} & {\scriptsize\bf
    AP1} & {\scriptsize\bf AP2} & {\scriptsize\bf AP1} & {\scriptsize\bf AP2} & {\scriptsize\bf AP1} & {\scriptsize\bf AP2} \\ 
  \hline \hline
  \multirow{5}{*}{\begin{sideways}\scriptsize\VisitProbInit\end{sideways}}
 & [0,1] & \textcolor{purple}{0.94} & \textcolor{blue}{0.99} & \textcolor{purple}{0.80} & \textcolor{blue}{0.89} & \textcolor{blue}{0.80} & \textcolor{purple}{0.42} & \textcolor{purple}{0.81} & \textcolor{blue}{0.99} & \textcolor{blue}{0.45} & \textcolor{purple}{0.13}\\  
& [0,7] & \textcolor{purple}{0.67} & \textcolor{blue}{0.99} & \textcolor{blue}{0.89} & \textcolor{purple}{0.88} & \textcolor{blue}{0.88} & \textcolor{purple}{0.48} & \textcolor{purple}{0.55} & \textcolor{blue}{0.99} & \textcolor{blue}{0.67} & \textcolor{purple}{0.21}\\ 
& [0,30] & \textcolor{purple}{0.59} & \textcolor{blue}{0.99} & \textcolor{purple}{0.91} & \textcolor{blue}{0.92}  & \textcolor{blue}{0.90}& \textcolor{purple}{0.56} & \textcolor{purple}{0.46} & \textcolor{blue}{0.98} & \textcolor{blue}{0.76} & \textcolor{purple}{0.29} \\  
& [30,60] & \textcolor{purple}{0.87} & \textcolor{blue}{0.99} & \textcolor{blue}{0.98} & \textcolor{purple}{0.31} & \textcolor{blue}{0.93} & \textcolor{purple}{0.00} & \textcolor{purple}{0.45} & \textcolor{blue}{0.96} & \textcolor{blue}{0.77} & \textcolor{purple}{0.00}\\ 
& [60,90] & \textcolor{purple}{0.91} & \textcolor{blue}{0.99} & \textcolor{blue}{0.97} & \textcolor{purple}{0.02} & \textcolor{blue}{0.96} & \textcolor{purple}{0.10} & \textcolor{purple}{0.56} & \textcolor{blue}{0.91} & \textcolor{blue}{0.83} & \textcolor{purple}{0.09} \\ 
 \hline
\multirow{5}{*}{\begin{sideways}\scriptsize\VisitCountInit\end{sideways}} 
& [0,1] & \textcolor{purple}{3.54} & \textcolor{blue}{14.58} & \textcolor{purple}{1.63} & \textcolor{blue}{2.24} & \textcolor{blue}{1.92} & \textcolor{purple}{0.72} & \textcolor{purple}{1.74} & \textcolor{blue}{5.77} & \textcolor{blue}{0.95} & \textcolor{purple}{0.28}\\ 
& [0,7] & \textcolor{purple}{1.19} & \textcolor{blue}{15.25} & \textcolor{blue}{2.21} & \textcolor{purple}{2.09} & \textcolor{blue}{2.75} & \textcolor{purple}{0.87} & \textcolor{purple}{0.84} & \textcolor{blue}{5.27} & \textcolor{blue}{2.11} & \textcolor{purple}{0.48} \\ 
& [0,30] & \textcolor{purple}{0.89} & \textcolor{blue}{15.55} & \textcolor{purple}{2.39} & \textcolor{blue}{2.52} & \textcolor{blue}{2.62} & \textcolor{purple}{1.19} & \textcolor{purple}{0.65} & \textcolor{blue}{4.75} & \textcolor{blue}{1.95} & \textcolor{purple}{0.69} \\
& [30,60] & \textcolor{purple}{2.28} & \textcolor{blue}{14.27} & \textcolor{blue}{5.06} & \textcolor{purple}{0.40} & \textcolor{blue}{4.29} & \textcolor{purple}{0.01} & \textcolor{purple}{0.80} & \textcolor{blue}{4.04}  & \textcolor{blue}{4.39} & \textcolor{purple}{0.01} \\
& [60,90] & \textcolor{purple}{3.00} & \textcolor{blue}{14.73} & \textcolor{blue}{4.48} & \textcolor{purple}{0.02} & \textcolor{blue}{4.61} & \textcolor{purple}{0.10} & \textcolor{purple}{1.28} & \textcolor{blue}{3.63}  & \textcolor{blue}{5.64} & \textcolor{purple}{0.82}\\ 
 \hline
\multirow{5}{*}{\begin{sideways}\scriptsize\StepCountInit\end{sideways}} 
& [0,1] & \textcolor{purple}{16.55} & \textcolor{blue}{4.53} & \textcolor{purple}{30.27} & \textcolor{blue}{22.75} & \textcolor{blue}{30.45} & \textcolor{purple}{90.36} & \textcolor{purple}{29.82} & \textcolor{blue}{12.01} & \textcolor{blue}{83.40}& \textcolor{purple}{332.40}\\ 
& [0,7] & \textcolor{purple}{44.55} & \textcolor{blue}{3.63} & \textcolor{blue}{22.14} & \textcolor{purple}{23.24} & \textcolor{blue}{23.27} & \textcolor{purple}{75.96} & \textcolor{purple}{63.24} & \textcolor{blue}{12.58} & \textcolor{blue}{45.55} & \textcolor{purple}{210.12} \\ 
& [0,30] & \textcolor{purple}{56.55} & \textcolor{blue}{3.44} & \textcolor{purple}{20.07} & \textcolor{blue}{19.28} & \textcolor{blue}{21.53} & \textcolor{purple}{59.94} & \textcolor{purple}{81.13} & \textcolor{blue}{13.68} & \textcolor{blue}{35.54} & \textcolor{purple}{145.35} \\
& [30,60] & \textcolor{purple}{23.41} & \textcolor{blue}{2.15} & \textcolor{blue}{9.02} & \textcolor{purple}{137.09} & \textcolor{blue}{18.90} & \textcolor{purple}{5483.99} & \textcolor{purple}{85.45} & \textcolor{blue}{15.97} & \textcolor{blue}{34.45} & \textcolor{purple}{25915.18} \\
& [60,90] & \textcolor{purple}{19.55} & \textcolor{blue}{2.23} &
\textcolor{blue}{10.46} & \textcolor{purple}{2269.78} &
\textcolor{blue}{15.49} & \textcolor{purple}{483.09} &
\textcolor{purple}{61.19} & \textcolor{blue}{21.01} & \textcolor{blue}{28.65} & \textcolor{purple}{532.74} \\ 
\hline
\end{tabular}
\end{table}

\begin{table}[!t] 
  \center
  \caption{AppTracker1. Characteristics of sessions in GPAM(2): \VisitProbInit\ for the state \UseStop\ computes the probability to reach the end of the session, while \SessionCount\ and \SessionLength\ compute the expected number of sessions and the expected number of time steps (button taps) in a session respectively.}
  \label{table:AT1_GPAM2_SAP_VP1_VC1_SC1_UseStop}
\begin{tabular}{|c|rr|rr|rr|}
  \hline
  {\scriptsize\bf Time} &
  \multicolumn{2}{c|}{\VisitProbInit} & \multicolumn{2}{c|}{\SessionCount} & \multicolumn{2}{c|}{\SessionLength} \\   
  \cline{2-3} \cline{4-5} \cline{6-7} {\scriptsize\bf interval} & {\scriptsize\bf AP1} & {\scriptsize\bf
    AP2} & {\scriptsize\bf AP1} & {\scriptsize\bf AP2} & {\scriptsize\bf AP1} & {\scriptsize\bf  AP2} \\  
\hline\hline
  $[0,1]$ & \textcolor{black}{0.99} & \textcolor{black}{0.31} & \textcolor{black}{10.13} & \textcolor{black}{0.37} & \textcolor{black}{3.86} & \textcolor{black}{130.96}
   \\ 
  $[0,7]$ & \textcolor{black}{0.99} & \textcolor{black}{0.43} & \textcolor{black}{10.20} & \textcolor{black}{0.54} & \textcolor{black}{3.81} & \textcolor{black}{87.76}
   \\ 
  $[0,30]$ & \textcolor{black}{0.99} & \textcolor{black}{0.38} & \textcolor{black}{10.82} & \textcolor{black}{0.47} & \textcolor{black}{3.51} & \textcolor{black}{102.07}
   \\ 
  $[30,60]$ & \textcolor{black}{0.99} & \textcolor{black}{0.99} & \textcolor{black}{6.17} & \textcolor{black}{7.55} & \textcolor{black}{7.09} & \textcolor{black}{5.36}
    \\ 
  $[60,90]$ & \textcolor{black}{0.99} & \textcolor{black}{0.99} & \textcolor{black}{5.43} & \textcolor{black}{7.35} & \textcolor{black}{8.28} & \textcolor{black}{5.56} \\
  \hline
\end{tabular}
\end{table}

\begin{table}[!t] 
  \center
  \caption{AppTracker1. GPAM(2). Inductive coding: categorisation of the average session counts and lengths, subsequently used for the 
  initial categorisation of activity patterns alongside the predominant states, and description of the usage intensities. 
  State identifiers (ids) are: 3 is \OverallUsage, 4 is
    \LastSevenDays, 5 is \SelectPeriod, 6 is \AppsInPeriod, 9 is
    \Stats, 10 is \UBCOverallUsage, 12 is \UBCStats, and 14 is  
    \UBCAppsInPeriod. State identifiers are highlighted as \colorbox{lgray}{\Summary\ states} and \colorbox{dgray}{\Specific\ states.} State 4 has a dual role as \Summary\ or \Specific\ depending on the context. 
    }
    \label{table:GPAM2_initial_AP_categ_AT1}
\begin{tabular}{c}
\begin{tabular}{|c|c|c|}
\hline
\multicolumn{3}{|c|}{\bf Categories of session counts} \\
\cline{1-3} {\bf few} & {\bf mid} & {\bf many}  \\
\hline
$\{0.37,0.47,0.54\}$ & $\{5.43,6.17,7.35,7.55\}$ &  $\{10.13,10.82\}$ \\
\hline
\end{tabular}
\\
\\
\begin{tabular}{|c|c|c|}
\hline
\multicolumn{3}{|c|}{\bf Categories of session lengths}\\
\cline{1-3} {\bf short} & {\bf mid} & {\bf long} \\
\hline
$\{3.51,3.81,3.86\}$ & $\{5.36,5.56,7.09,8.28\}$ & $\{87.76,102.07,130.96\}$ \\ 
\hline
\end{tabular}
\\
\\
\begin{tabular}{|c|cc|cc|}
\hline
{\bf Time} & \multicolumn{2}{c|}{\bf AP1} & \multicolumn{2}{c|}{\bf AP2} \\
\cline{2-3} \cline{4-5} {\bf interval} & {\bf Sessions} & {\bf Predominant}
& {\bf Session} & {\bf Predominant}\\ 
& {\bf characteristics} & {\bf states}
& {\bf characteristics} & {\bf states}\\ 
\hline\hline
$[0,1]$ & many/short & \colorbox{dgray}{5},\colorbox{lgray}{9},\colorbox{dgray}{4} & few/long & \colorbox{lgray}{3},\colorbox{lgray}{9},\colorbox{lgray}{4},\colorbox{lgray}{10},\colorbox{lgray}{12}\\
\hline
$[0,7]$ & many/short & \colorbox{dgray}{5},\colorbox{dgray}{4},\colorbox{dgray}{6} & few/long &  \colorbox{lgray}{3},\colorbox{lgray}{9},\colorbox{lgray}{4} \\
\hline
$[0,30]$ & many/short & \colorbox{dgray}{5},\colorbox{dgray}{4},\colorbox{dgray}{6}  & few/long &  \colorbox{lgray}{3},\colorbox{lgray}{9},\colorbox{lgray}{4},\colorbox{lgray}{12}\\
\hline
$[30,60]$ & mid/mid & \colorbox{dgray}{4},\colorbox{dgray}{5},\colorbox{dgray}{6},\colorbox{dgray}{14} & mid/mid &  \colorbox{lgray}{3},\colorbox{lgray}{9},\colorbox{lgray}{10},\colorbox{lgray}{12} \\
\hline
$[60,90]$ & mid/mid & \colorbox{dgray}{6},\colorbox{dgray}{5},\colorbox{dgray}{4},\colorbox{dgray}{14} & mid/mid &  \colorbox{lgray}{3},\colorbox{lgray}{10},\colorbox{lgray}{9},\colorbox{lgray}{12} \\
\hline
\end{tabular}
\\
\\
\begin{tabularx}{0.8\linewidth}{rX}
\toprule
{\bf Usage intensity} & {\bf Usage intensity characteristics}\\
\midrule
\Browsing &  few/long sessions centred on at least three 
  states in the \TopLevelMenu, sometimes from different state groups\\ 
\Glancing &  many/short sessions centred on one or two states\\
\Focussing & mid/mid sessions centred on states from the same group \\
\bottomrule
\end{tabularx}
\end{tabular}
\\
\bigskip
\begin{tabular}{ccll}
\toprule
{\bf Time interval} && {\bf AP1 label} & {\bf AP2 label}   \\
\midrule
$[0,1]$ && \Glancing\Specific & \Browsing\Summary \\
$[0,7]$ && \Glancing\Specific & \Browsing\Summary \\
$[0,30]$ && \Glancing\Specific & \Browsing\Summary \\
$[30,60]$ && \Focussing\Specific & \Focussing\Summary \\
$[60,90]$ && \Focussing\Specific & \Focussing\Summary \\
\bottomrule
\end{tabular}
\end{table} 

\begin{figure*}[!t]
  \centering 
  \subfigure[AppTracker1]{\includegraphics[width=0.48\textwidth]{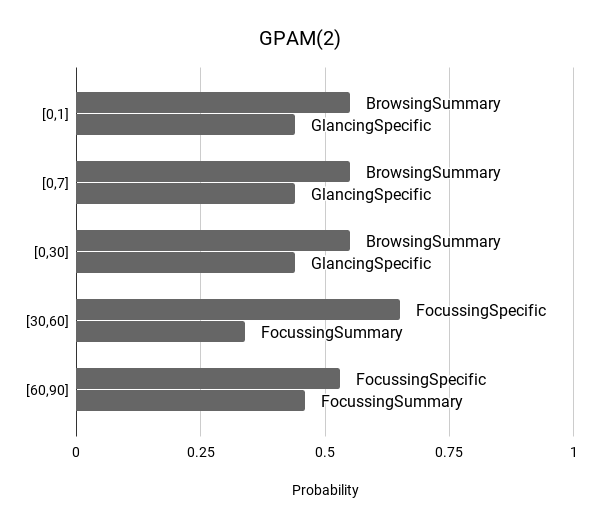}\label{fig:chart_longrun_GPAM2_AT1}}
\quad 
  \subfigure[AppTracker2]{\includegraphics[width=0.48\textwidth]{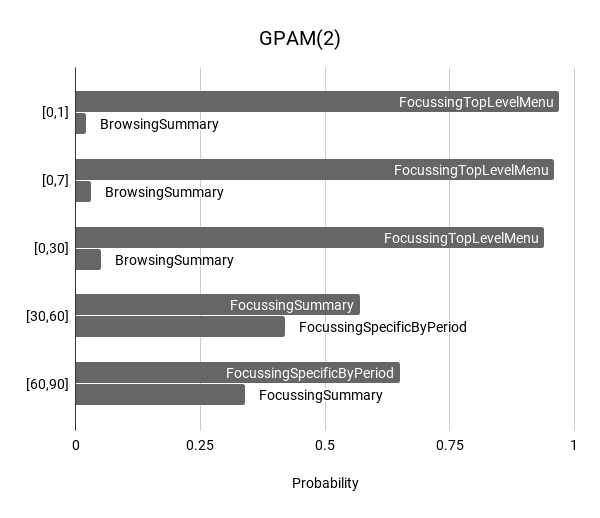}\label{fig:chart_longrun_GPAM2_AT2}}

 \caption{AppTracker1 and AppTracker2. Visualisation of probabilities of being in each
    activity pattern in the long run in GPAM(2)  models for
    different time intervals; the entries are ordered decreasingly for each time interval. These results are based on model checking the temporal property \LongRunPattern\ for each activity pattern.  
    }
\label{fig:at1_at2_gpam2_longrun}
\end{figure*}

\smallskip

An important question to ask is: {\em  does the latent structure
we uncover simply reflect the top level  menu structure?}  To answer it we looked at GPAM(3) models (see Appendix~\ref{appendix_gpam3}), since there are three main menu options (excluding \Settings). 
If the activity patterns simply reflect the menu structure, then when
  we would expect each of the patterns in GPAM(3) to be centred on one the
above states, with very low correlations between pairs of those
states. This was not the case.  
In all five GPAM(3) models we
found activity patterns with either \Focussing\ or \Glancing\ usage
intensity, centred around \LastSevenDays\ and either \SelectPeriod\ or
\AppsInPeriod; there was no model with one pattern centred on either
\SelectPeriod\ or \AppsInPeriod\ but no \LastSevenDays\, and another
pattern centred on \LastSevenDays, but not on \SelectPeriod\ or
\AppsInPeriod. 

\subsection{Redesigning the Interface for Experienced Usage}

It became evident that AppTracker1's   top-level
menu was not a good fit for   users' interaction styles. For example, we uncovered \Glancing\
 behaviours, where users would quickly consult the app to view their
usage behaviour,  but this did not fit well with the  menu
structure: a user could glance at all-time most-used apps
(\OverallUsage\ state) within the first menu item, but if they had
used AppTracker1 for a long time, it becomes increasingly likely that
this list would be static.  Conversely, while a user could find recent
(e.g.~today's) usage, this would involve several steps through the
menu of more detailed information.

We
chose to concentrate  on redesign for experienced usage; these were users who  had  voluntarily continued with the app over a period of time 
 so  we also expected to 
see more stable interaction styles, where initial user learning processes have subsided. 
An  equally valid alternative  would  have 
been early stage users, for example to work on issues that may improve retention of 
users beyond the initial experiences.

  Our aim was not to change the overall purpose of
AppTracker, or to add new features, but   to   reconfigure the  menu structure, by adding, removing or
moving states, to   increase the support
 for more  efficient \Glancing\ on \Summary\ states and \Focussing\
on \Specific\ states.
These
behaviours  were   not well aligned with the existing
menu layout.  For example, we see \AppsInPeriod\ (from the \Specific\
sub-menu) appear in \Glancing\ activity patterns, because users might
drill down into the specific menus to see the current day's
activity. Table~\ref{table:GPAM2_initial_AP_categ_AT1} also shows
activity clustering around \Summary\ and \Specific\ states during
experienced usage, but not in an efficient way; users are seen to be
\Focussing\ on \Summary\ states, when such information could be
presented in such a way as to allow glancing to acquire the desired
information more quickly.

As a consequence we changed the top level menu structure to offer two
options (plus \Settings)  instead of three. The two options correspond
to \Glancing\ and \Browsing\ usage, and to \Focussing\ usage, and we
call them {\em My Top Apps} and {\em Explore Data}, respectively. The
corresponding two state groups are \Summary\ states and \Specific\
states (as the union of \SpecificMenu, \SpecificByPeriod,
\SpecificByApp, and \SpecificStats\ state groups).
Hence,
several states further down the hierarchy are moved or split,
i.e.~moving states upon which an activity pattern is centred to be
close to the option associated with that activity pattern.
To support \Glancing\ and \Browsing\ usage behaviour, {\em My Top
  Apps} contains only tables showing: (i) the user's most-used apps
since installation of AppTracker and (ii) the most-used apps on the
current day. This way the redesigned menu aims to make today's usage
much more easily accessible from the top-level menu.

\begin{figure*}[!t]
  \centering 
  \includegraphics[width=0.95\textwidth]{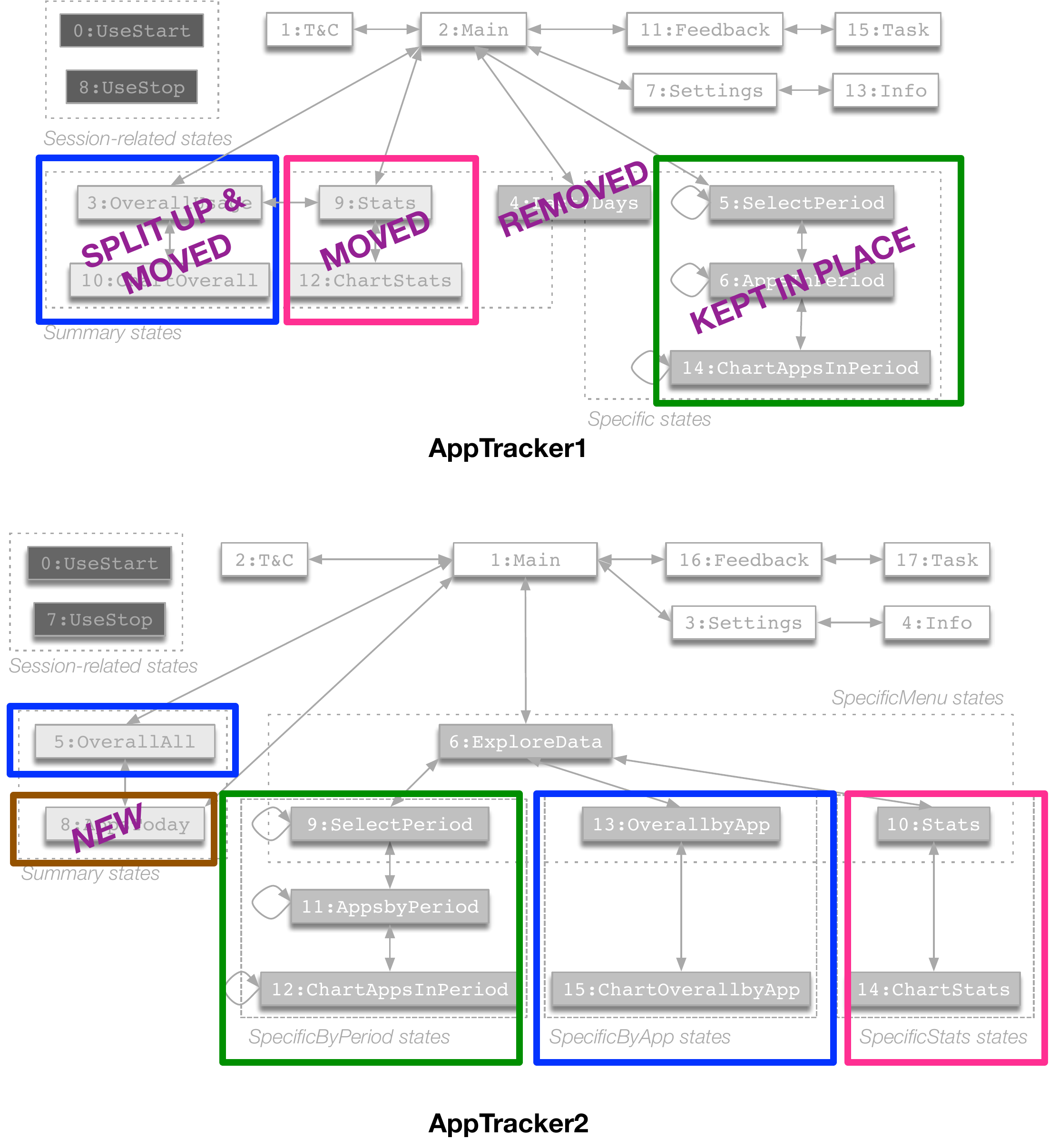}
  \caption{AppTracker1  and AppTracker1 interface designs.  We use same colours between
    the two app versions to show correspondence.}
\label{fig:redesign}
\end{figure*}

The changes are     illustrated in Fig.~\ref{fig:redesign}  and summarised as:
\begin{itemize}

\item The app's menu is restructured to specifically
  support as two main styles of use \Glancing\ and \Focussing, and
  therefore to only have two main menu options to make these behaviours
  more distinct.


\item The screen view \OverallUsage\ was replaced by
  a new, more glancing-like view \OverallAll\, that does not allow for
  drilling down into detailed usage.


\item A new glancing-like screen view \AppsToday\ is
  included in the \Summary\ part of the menu.


\item The screen view \LastSevenDays\ was removed. 


\item The \Specific\ states group of the menu was
  broken up into more subtle sub-components: \SpecificByPeriod,
  \SpecificByApp, \SpecificStats.


\item The screen view \OverallUsage\ was moved into
  the \Specific\ part and renamed as \OverallbyApp\ alongside with
  \UBCOverallbyApp; these two new states are grouped into
  \SpecificByApp.


\item The screen view \Stats\ was moved into the
  \Specific\ part of the menu alongside with \UBCStats; these two new
  states are grouped into \SpecificStats.
\end{itemize}

The main menu screen of AppTracker2 offers three main options
(Fig.~\ref{fig:AT2menu}), and there are 18
user-initiated events. New  states are:


\begin{figure*}[!t]
  \centering 
  \subfigure[Main Menu View]{\includegraphics[scale=0.16]{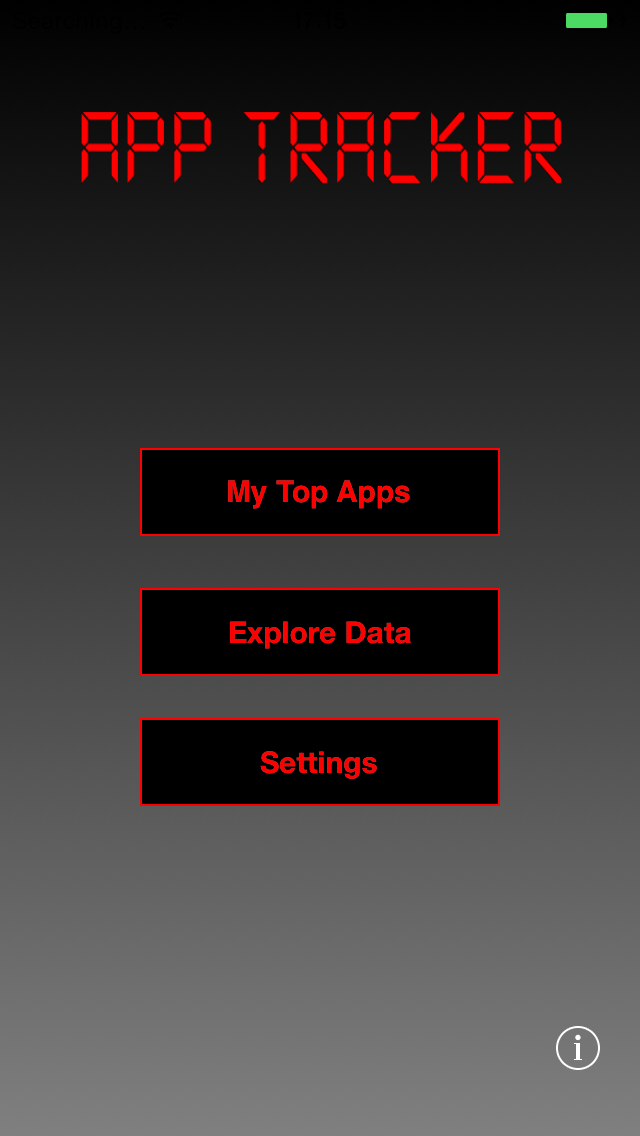}\label{fig:AT2menu}}
  \quad
  \subfigure[My Top Apps Sub-menu]{\includegraphics[scale=0.16]{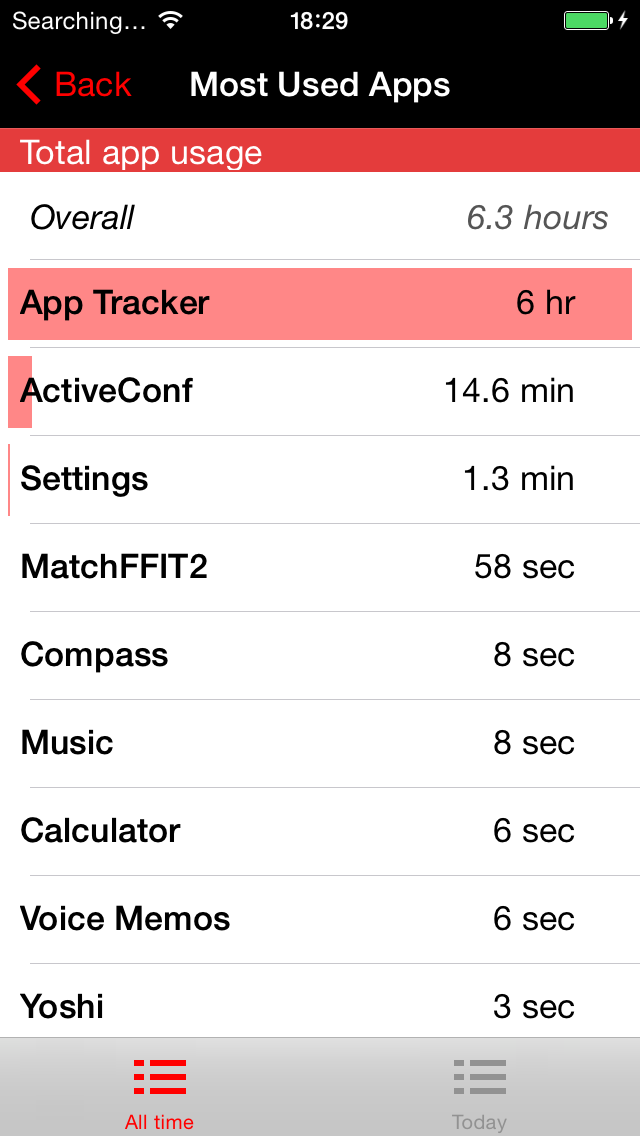}\label{fig:TopApps}}
  \quad  
  \subfigure[Explore Data Sub-menu]{\includegraphics[scale=0.16]{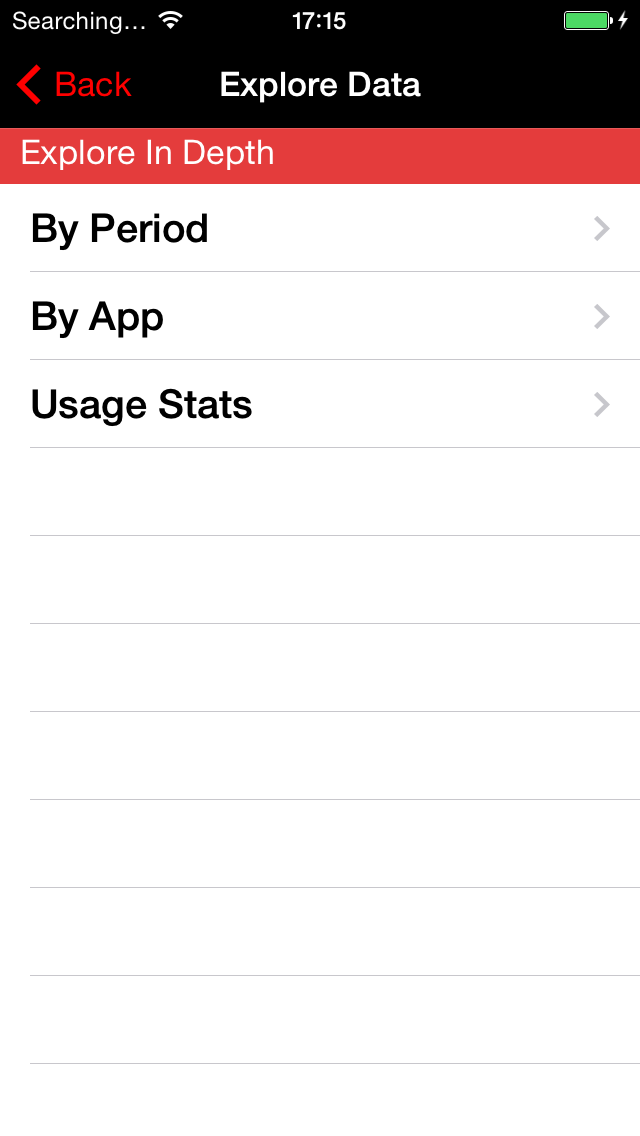}\label{fig:AT2explore}} 
  \caption{Screenshots from AppTracker2: (a) The main menu view
    corresponds to the state \Main; the {\em My Top Apps} option takes
    the user to the state \OverallAll, {\em Explore Data} to the state
    \ExploreData, and {\em Settings} to the state \Settings. (b) The
    {\em My Top Apps} view (corresponding to the state \OverallAll)
    shows the overall usage of the device since installing the app and
    offers the possibility to view today's usage -- the bottom-right
    option {\em Today}. (c) The {\em Explore Data} view offers the
    options of exploring the data {\em By Period} (state
    \SelectPeriod), {\em By App} (state \OverallbyApp), and viewing
    the {\em Usage Stats} (state \Stats). }
\label{fig:screenshotsAT2}
\end{figure*}

\begin{itemize}[label=$\bullet$]
\item \OverallAll: summary statistics about the overall device usage
 since installing AppTracker2,
\item \ExploreData: three options for more in-depth exploration of all
  recorded data,
\item \AppsToday: summaries of the current day's usage of apps,
\item \AppsbyPeriod: usage statistics for various apps by selected time period, 
\item \UBCAppsInPeriod: detailed app usage when selected from \AppsbyPeriod,
\item \OverallbyApp: usage summary for a selected app,
\item \UBCOverallbyApp: detailed app usage, when selected from \OverallbyApp.
\end{itemize}

States are grouped,  depending on  position in
the overall AppTracker2 menu, insights  gained from the interaction styles
of AppTracker1, and design intentions of AppTracker2.
AppTracker2 was released in 2016 and our data sets are taken from a
sample of 600 user traces over a period of six months.

\subsection{AppTracker2 Interaction Styles}

For brevity we show only our working for the inductive coding process for GPAM(2) models in  Table~\ref{table:AT2_APlabels_GPAM2}. The labels are self-explanatory, but it is important to note that \Summary\ states are a different grouping than in AppTracker1,  
 see Fig.~\ref{fig:redesign}.  In particular the new state \AppsToday\ (state 8) features in all the activity patterns except \Focussing\SpecificByPeriod.
The main result is  again  there are four distinct activity patterns and  a clear split between styles for early days usage
and experienced usage. 
Long run probabilities  are  in Fig.~\ref{fig:steadystate_AT1_AT2}(b) and probabilities that for a given activity
pattern, a state leads to activity pattern change (within a session), 
are given in Fig.~\ref{fig:at1_at2_gpam2_state2pattern}(b).

The interaction styles for GPAM(2) models of AppTracker2 usage are summarised as follows. 
Overall, there are fewer and shorter sessions, there are no short
sessions and no mid/mid   combinations. There is much more \Focussing\  behaviour.
There is very little \Browsing\ in the early days models, which could
indicate that AppTracker2 is easy to use, or that most of the users
are already experts in using this second version of the app. 
The probabilities to transition are low across all models.

\begin{table}[!t]
  \center
  \caption{AppTracker2, GPAM(2) models. Categorisation of the average session counts and lengths, subsequently used for the 
  initial categorisation of activity patterns alongside the predominant states, and description of the usage intensity labels. 
  State identifiers (ids) are: 5 is \OverallAll, 6 is
    \ExploreData, 8 is \AppsToday, 9 is \SelectPeriod, 10 is
    \Stats, 11 is \AppsbyPeriod, 12 is \UBCAppsInPeriod, 13 is \OverallbyApp, 14 is \UBCStats, 15 is \UBCOverallbyApp. State identifiers are highlighted as \colorbox{lgray}{\Summary\ states} and \colorbox{dgray}{\Specific\ states.}}  
\label{table:AT2_APlabels_GPAM2}
\begin{tabular}{c}
\begin{tabular}{|c|c|}
\hline
\multicolumn{2}{|c|}{\bf Categories of session counts} \\
\cline{1-2} {\bf few} & {\bf many}  \\
\hline
$\{1.40, 2.10, 3.04\}$ & $\{5.35, 5.86, 5.87, 6.23, 6.39, 6.52, 7.02\}$  \\
\hline
\end{tabular}
\\
\\
\begin{tabular}{|c|c|}
\hline
\multicolumn{2}{|c|}{\bf Categories of session lengths}\\
\cline{1-2} {\bf mid} & {\bf long} \\
\hline
$\{6.14, 6.57, 6.68, 6.88, 7.52, 7.84, 8.31\}$ & $\{17.61, 25.08, 38.84\}$  \\ 
\hline
\end{tabular}
\\
\\
\begin{tabular}{|c|cc|cc|}
\hline
{\bf Time} & \multicolumn{2}{c|}{\bf AP1} & \multicolumn{2}{c|}{\bf AP2} \\
\cline{2-3} \cline{4-5} {\bf int.} & {\bf Sessions} & {\bf States ids}
& {\bf Sessions} & {\bf States ids}\\ 
\hline\hline
$[0,1]$ & few/long & \colorbox{lgray}{5,\ 8},\colorbox{dgray}{6} & many/mid & \colorbox{lgray}{5},\colorbox{dgray}{6},\colorbox{lgray}{8},\colorbox{dgray}{9} \\
\hline
$[0,7]$  & few/long & \colorbox{lgray}{5,\ 8},\colorbox{dgray}{6} 
& many/mid & \colorbox{lgray}{5},\colorbox{dgray}{6},\colorbox{lgray}{8},\colorbox{dgray}{9}\\
\hline
$[0,30]$ 
& few/long & \colorbox{lgray}{5},\colorbox{dgray}{6},\colorbox{lgray}{8} 
& many/mid & \colorbox{lgray}{5},\colorbox{dgray}{6},\colorbox{lgray}{8},\colorbox{dgray}{9} \\
\hline
$[30,60]$ 
& many/mid & \colorbox{dgray}{9,\ 6,\ 11,\ 12}
& many/mid & \colorbox{lgray}{5,\ 8,\ 6},\colorbox{dgray}{6} \\
\hline
$[60,90]$ 
& many/mid & \colorbox{dgray}{6,\ 9,\ 11,\ 12} 
& many/mid & \colorbox{lgray}{5,\ 8},\colorbox{dgray}{6} \\
\hline
\end{tabular}
\\
\\
\begin{tabularx}{0.7\linewidth}{@{}rX@{}}
\toprule
{\bf Usage intensity} & {\bf Usage intensity characteristics}\\
\midrule
\Browsing & few/long sessions centred on at least three 
  states in the \TopLevelMenu, sometimes from different state groups\\ 
\Focussing &  many/mid  sessions centred on states from the same group \\
\bottomrule
\end{tabularx}
\\
\\
\begin{tabular}{@{}cll@{}}
\toprule
{\bf Time int.} & {\bf AP1 label} & {\bf AP2 label}   \\
\midrule
$[0,1]$ & \Browsing\Summary & \Focussing\TopLevelMenu  \\
$[0,7]$ & \Browsing\MyTopApps & \Focussing\TopLevelMenu \\
$[0,30]$ & \Browsing\MyTopApps & \Focussing\TopLevelMenu \\
$[30,60]$ & \Focussing\SpecificByPeriod & \Focussing\MyTopApps \\
$[60,90]$ & \Focussing\SpecificByPeriod & \Focussing\MyTopApps \\
\bottomrule
\end{tabular}
\end{tabular}
\end{table}

\subsection{Comparing Interaction Styles in AppTracker1 and AppTracker2}

Comparing the interaction styles in AppTracker2 with AppTracker1, there are many similarities between activity patterns, and these are reflected in the label names.     This is not surprising, given the redesign involved menu 
re-organisation, not a full-scale application redesign. The major differences between the styles can be seen in Fig.~\ref{fig:at1_at2_gpam2_longrun} and 
Fig.~\ref{fig:at1_at2_gpam2_state2pattern}. We summarise  the {\em effects} of the
redesign  on interaction styles as: AppTracker2 has  
      no short sessions,
    more \Focussing\ in all    models, with \Focussing\ on  specific states more likely for experienced users, 
 very little \Browsing, and with low likelihood, and only in
 early days usage models,   
extensive occurrences of the  new summary  state \AppsToday, and  
   a  lower likelihood to move out of an activity pattern. The last may
  indicate more alignment with user intention and in particular with experienced user, or  that most of the users
are already experts in using this second version of the app. One could argue that the \Focussing\TopLevelMenu\ observed in the early days usage is a form of longer,  but focussed \Glancing. It is interesting to note that while the redesign was targeted mainly at experienced users, it had a significant,  positive effect on early days users.   


\subsection{Insights from Models with at Least Three Clusters} 
Increasing the number of clusters $K$ to $3$ revealed 
more fine-grained interest in selected states or sets of states,  and also the possibility of  very short sessions  in AppTracker2. For example,  in GPAM(3) we uncovered  \Glancing\  again  for experienced usage, but it was entirely focussed  on the Main menu or on the (new) \Summary\ states.  A brief analysis of GPAM(3) models for both AppTracker1 and AppTracker2 can be found in the Appendix~\ref{appendix_gpam3}. 

We observed that as we increase the number of model components, we see
some finer-grained interest in selected states or groups of states
(e.g.~in \Focussing\Specific\ in GPAM(4) and GPAM(5) models) that was
less prominent for smaller values of $K$. For $K \geq 4$ we see more
activity patterns that are centred around each of the
states. There is no optimal value
for $K$, in general. It is an exploratory tool whose  usefulness  
 depends on the number of state groups, the state diagram,  and the granularity that is helpful for redesign. In this study we found 2-3 clusters most useful.  Note, we needed to study at least $K=3$ to investigate
whether activity patterns align to top level menu choices. It was also
important to analyse GPAM(4) and GPAM(5) models to confirm they did
not reveal significantly different or more useful (for redesign)
activity patterns; details of GPAM(4) and GPAM(5) analyses are not included here.



%% file: discussion.tex
We reflect on some implications of working with data-driven ML models
and computational methods.

\subsection{Data Reliability and Segmentation}

We identified and subsequently fixed several errors in the logs such
as missing start or end of sessions, missing events, unexpected
timestamps, all due to the interactions between the logging framework
SGLog and iOS such as apps crashing before data were written to file,
or network failures in transmitting data to our servers for analysis.
We discarded the user traces with less than five sessions because our
focus was on studying long term engagement and such short user traces
would only add more noise in the data set.  The choice of five
sessions as cut off for the minimum user trace length was empirical
and we cannot exclude that a different minimum would produce different
results.

We selected different time intervals that covered experiences from
initial engagement (first day) to extended engagement (over several
months).  These intervals were of interest, and made sense to us, for
this application.  More generally, designers might choose to focus on
different intervals, at different times. For example, shortly after
release of an initial design, they may focus on models inferred from
first day usage (or even shorter, e.g.~the first five minutes), but in
subsequent designs they may choose to consider only models inferred
from usage data from users that engage for long periods of time.

\subsection{Bias in Data Sets} 

AppTracker2 was released as an update to AppTracker1 (both available
only on the Cydia app store for jailbroken iOS devices).  As such, our
set of AppTracker2 users could be either existing users who installed
the update, or new users coming to the app afresh. This might have an
effect on interaction styles, because existing users would be more
familiar with the system, and would also have existing logged
information recorded, so would be looking at charts and tables already
populated with data, whereas new users would see empty screens during
the early days.  More generally, it can be a challenge to distinguish
new users, because users can reinstall an app or purchase new
hardware. For example, Apple altered the type of unique device
identifiers that apps can access, to track users across installations.
More recently Apple have prohibited this entirely, such that if a user
uninstalls all apps from a developer and then re-installs, an entirely
new identifier is generated, and nothing is provided that can link the
user of the second installation to the first one (and vice
versa)~\cite{BojinovMNB14,RooksbyAMMGC16}.  These are issues inherent
in performing real world deployments of apps rather than conducting
more constrained lab-style research. We would argue that great
benefits are gained from the large number of users we were able to
recruit, and the external validity gained through users
`self-selecting' to download our app rather than having being
explicitly recruited as participants in a trial, and our subsequent
study of people's use of our apps on their own devices embedded in
their everyday lives.

\subsection{Selection of States, Volume of Data, and Scalability}
Our study gained from having significant volume of log data to work
on, for reliable application of ML methods. It is important to note
that neither the volume of log data nor number of user traces is
relevant to probabilistic model checking; only the number of observed
states determines its complexity.  The AppTracker study involved a
vocabulary of 16-18 observed states, which was comfortably manageable,
but model-checking would have difficulty with an order of magnitude
larger. This could arise in an app with much richer menu options, or
further inclusion of categorical variables such as location,
demographic information, duration of the event, time of day or week,
etc.

 \subsection{User Proficiency and Focus over Time}
  Simpler forms of analytics may chase absolute
 metrics such as spending longer in the app or launching the app more
 often as measures of engagement~\cite{AttfieldKLP11}.  We did we not
 conflate (subject) intensity of usage with usage expertise and/or
 proficiency. We aimed to support the styles that were observed, and
 note that both \Focussing\ and \Glancing\ styles were present in
 experienced usage models.

\subsection{Further Redesign Possibilities } 
\noindent 
{\bf New users. }
  If there is a difference between new and longer-term users (e.g. first day/week/month
users vs. second/third month users),   then add new functionality that supports   application ‘onboarding’ and a smooth transition into another style of  interaction.

\noindent 
{\bf Micro-usages.}
Look for activity carried out in very brief micro-usages that might be better served by widgets rather than navigating the full application. By widget we mean a simple additional element of a
device’s graphical interface, which is usually complementary to an application running on that device.

\noindent 
{\bf Shortcuts.}
 Identify the most  popular initial states in an activity pattern and implement shortcuts to these  states.   

\noindent 
{\bf Split the application.} 
If  there is (nearly) always only one activity pattern per session  and patterns do not overlap, then split the application into two (or more) separate applications.

%% file: related.tex
\subsection{Our Previous Work}\label{previouswork}

 We developed our analytics over a number of years:  refining the Markovian models, the temporal properties, and segmentation of data sets.
 Our first study~\cite{AndreiCHG14} involved the iOS multi-player game
Hungry Yoshi \cite{McMillanMBHC10}, one data set,   a simple inferred model, simple temporal logic properties, and $K = 2$.   The app  was different  to AppTracker in that it had clear (user) goal
states and (in addition to user interactions)  external events when the device picked up  scanned Wi-Fi access points. 
   We uncovered two interaction styles   representing different strategies for playing the game, but    could not   implement a  redesign because Apple's iOS changes    meant    we could no longer   scan for Wi-Fi access points. The app is no longer under development.  Our initial analysis of AppTracker~\cite{AndreiCCMR16} used a different  inferred  admixture   model,   a restricted set of generic  temporal  properties, and segmented data sets. There was no inductive coding nor menu redesign. The GPAM model was   defined in~\cite{AndreiC18}, where we also  introduced the possibility of logic formulae over the latent variables, though we did not consider   the probability to transition between activity patterns. Significantly, none of our earlier work involved  specific design recommendations, the implementation and deployment of a new design, nor   analysis of interaction styles in   the new design.

\subsection{Other Related Work}

Our approach to modelling 
was motivated initially by an
empirical study of simplicial mixtures for modelling webpage browsing
and telephone usage~\cite{GirolamiK04}.  We note that in the same
year, Bowring et al.~\cite{BowringRH04} referred to Girolami et al.'s
work~\cite{GirolamiK04} when suggesting that a hidden Markov model for
automatic classification of software was possible future work.  To our
knowledge, no one else has investigated admixture models for modelling
interaction behaviour, however the existence of different
user-populations or user types among the larger mobile app user
population was also recently highlighted
in~\cite{ChurchFBL15,JonesFHGK15,ZhaoR16} and mixture models proposed.

Markov chains as models for software usage were proposed nearly 30
years ago by Whittaker and Poore~\cite{WhittakerP93}, where transition
probabilities are estimated from the frequency counts of all bigrams
occurring in the execution traces. We can consider this model as a
GPAM with only one component, i.e.~GPAM(1).  Markov chain models of
software behaviour have also been employed
in~\cite{CookW98,ThimblebyCJ01} and more recently
in~\cite{GhezziPST14,BusanyM16,LuckowP16,KostakosFGH16,Wang0YP17,EmamM18}.
Usage styles are uncovered using statistical methods
in~\cite{KostakosFGH16}, though not as latent states in a hidden Markov model
variant. The analysis techniques used in~\cite{WhittakerP93} or
in~\cite{ThimblebyCJ01} are classic mathematical operations on the
transition matrix such as computing the long run probability of being
in one state or the expected number of states transitions to first
reach a state (mean first passage time), whereas we use probabilistic
temporal logic properties, which allows for more expressive properties
to be formulated and then analysed automatically in a probabilistic
model checking tool.  First-order Markov models have been used for
modelling in many related problems such as: human navigation on the
Web, where states correspond to visited
webpages~\cite{BorgesL00,ChierichettiKRS12,SingerHTS14,SingerHHS15,GhezziPST14},
more specifically clickstream
data~\cite{montgomery2004modeling,LuDM05,BenevenutoRCA09}, usability
analysis, where states correspond to button presses in
general~\cite{ThimblebyCJ01}, mobile applications, where states
correspond to device screen events~\cite{AndreiCHG14,KostakosFGH16},
and human interactions with a search engine~\cite{TranMFA17}.

Research on mobile app usage by Banovic et al.~\cite{BanovicBMD14}
presents evidence for (and characterises sessions based on) duration
and interaction types such as glance, review and engage. We also
identified three types of usage intensity (\Glancing, \Browsing\ and
\Focussing), however they are characterised by the number of
in-session interactions and frequency of sessions.  We note that both
our \Glancing\ and the glance of~\cite{BanovicBMD14} involve
micro-usages.  We suggest our characterisation of glancing behaviour
is closer to that of checking habit~\cite{OulasvirtaRMR12} as brief,
repetitive inspection of dynamic content quickly accessible on the
device.  The BEAR tool~\cite{GhezziPST14} is similar in that it infers
discrete time Markov chains from logs and probabilistic temporal logic
properties and PRISM are used to query the models; but the key
difference is users are classified according to static
attributes,~e.g. by time zone or operating system, which the designer
has pre-defined, and behaviours are assumed to be
homogeneous,~i.e. there is no in-class variation and no ability to
express or detect hybrid behaviours.
This issue is also raised in~\cite{BusanyM16}, arguing that the
filters for partitioning the log data can have a dramatic effect on
the resulting model and the subsequent analyses.

We also note DTMCs are models for usage patterns in~\cite{Stragier19},
where mHealth apps are analysed based on visualisation of the
interactive graphical representation of DTMCs and clustered sequences
of various lengths.  As mentioned above, probabilistic temporal logic
provides an additional analytical tool for analysing DTMCs beyond
insights gained from graphical representations.

Other computational interaction approaches to modelling human
behaviour, leveraging logs of user-initiated events, include Markov
decision processes (MDPs)~\cite{BanovicBCMD16} (identifying routines
across individuals and populations) and partially observable Markov
decision processes (POMDPs)~\cite{ChenBBOH15,HowesCAL18}. We could
analyse properties of such Markovian models with probabilistic model
checking, and thus bring into play the power of temporal logic and
interpretation of results following our approach.


%% file: conclusions.tex

Interaction redesign and data go hand in glove, but previously we did
not have the quantitative tools to uncover styles that are more
nuanced than tasks, especially at a large scale.  In this paper we have shown how  new computational methods, based on unsupervised  ML clustering and probabilistic temporal logic, provide  such a new quantitative tool for studying and interpreting interaction styles.     

The admixture Markov
models we inferred embody the ways user wanted to, and actually {\em
did}, use AppTracker.  The study results were a revelation to us: we
had no preconceptions about possible differences between early days
and experienced usage, and what kind of activity patterns we would
find in both AppTracker1 and AppTracker2.  We found that AppTracker1’s
top-level menu was not a good fit for the ways that users interacted
with the app. For example, we identified experienced usage styles
consisting mainly of \Glancing\ and \Focussing\ patterns that are
centred on AppTracker1's \Summary\ and \Specific\ states
respectively. These styles were not aligned with the existing menu
layout.  For example,
\AppsInPeriod\ (from the \Specific\ sub-menu) appears in \Glancing\
activity patterns, because users drill down into the specific menus to
see the current day's activity. And conversely, users are seen to be
\Focussing\ on \Summary\ states, when the information could be
presented in such a way as to allow it to be acquired in a \Glancing\
behaviour.  Consequently we re-designed the interface, offering only
two main options instead of three, and moving states upon which an
activity pattern is centred to be close to the option.  The
interaction styles we uncovered in AppTracker2 showed that it supports
the purpose of the redesign, with no short sessions, more \Focussing\
in all models, and \Browsing, which was prevalent for early days usage
in AppTracker1, almost disappeared and had only a low likelihood in
early days usage models.  Overall the likelihood to transition between
activity patterns was reduced, indicating that users more quickly
found a suitable style.  These insights were only possible through the
use of admixture (as opposed to mixture) Markov models and
the \StateToPattern\ property that includes latent variables.

Many, if not all, of the general concerns about use of ML and  computational methods applied to
our study of interaction design include bias in data, data reliability, and
temporal validity of model. The first two are a consequence of
real-world deployment, rather than lab-based studies.  The first
included a bias we could not eliminate: the possible effects of
existing users on the second design.  The inability to distinguish
existing and new users with absolute certainty is an aspect of
real-world app deployments, and subsequently for the use of
computational methods.  The second is also an aspect of real-world
deployments as interactions between the different systems, including
those for communications and data collection, affect data reliability.
The last is pertinent to redesign -- how often to infer models from
new data sets depends on how quickly and to what degree the underlying
data is changing. 
It is sometimes tempting, when applying computational methods, to
refer to "the data set" as if it were one monolithic entity.  This
study has highlighted the impact of data segmentation (in our case,
temporal segmentation) on the models and subsequent longitudinal 
analysis and decisions.  

 We emphasise that data never actually speaks for
itself, it is up to the analyst to pose meaningful questions and
visualisations. A crucial tool for the analyst in posing these questions is    probabilistic temporal
logic formulae over the latent variables - the activity patterns uncovered in the ML inference. 
We require a {\em temporal} logic to reason about computation paths,  to express relationships between observed states and behaviours within a session.  We found the properties concerning likelihood of changing activity patterns (\StateToAPProb) and long run behaviour (\LongRunPattern) to be the most powerful and useful aspects of our analytics, as these allowed us to see which activity patterns were more popular or transient, for given time intervals, hence gaining additional insight into different interaction styles to inform redesign.

As future work, we can add categories of dwelling time before each
 event and static user attributes.  We could also introduce
 qualitative methods such as user interviews, to gain insight into the
 intent behind an activity pattern. A novel, on-line approach to this
 would be to pop-up a questionnaire when a user employs a specific
 activity pattern for a length of time, or starts to employ to a
 specific pattern; this would extend context-triggered experience
 sampling methods proposed in~\cite{IntilleRKAB03}. A complementary, more passive,
 on-line approach would be to fire up, selectively, more frequent,
 fine-grained and/or different tracking and sensing so as to gain a
 more in-depth picture, in a temporary way that is mindful of the fact
 there may be costs to users such as increased battery drain and data
 transmission charges. We could also experiment with data sets from
 different sub-populations, e.g. to compare activity patterns of users
 that engage for a long time, with those who disengage after as short
 period of time.  Finally, we could automate the visualisations and
 allow interactions with the analytics, e.g.~click on an individual bar in long run
 probability charts, which would show corresponding predominant states.

%% file: appendix_AppTracker.tex
\section{AppTracker States}\label{lab:AppTrackerStates}

The states in AppTracker 1 are the following:
\begin{itemize}
\item \Summary\ states:
\begin{itemize} 
\item \OverallUsage: summary of all recorded data,
\item \Stats:  statistics of app use, 
\item \UBCOverallUsage: detailed app use, when selected from \OverallUsage,
\item \UBCStats: detailed app use, when selected from \Stats, 
\item \LastSevenDays: last seven days of top five apps used, 
\end{itemize}

\item \Specific\ states:
\begin{itemize} 
\item \SelectPeriod: statistics for a selected time period,
\item \AppsInPeriod: apps used for a selected time period,
\item \UsageBarChartAppsInPeriod: detailed app use, when selected from
  \AppsInPeriod,
\item \LastSevenDays: last seven days of top five apps used, 
\end{itemize}

\item \Session-related\ states:
\begin{itemize} 
\item \UseStart: start of a session (launch or bring AppTracker1 to
  the foreground),
\item \UseStop: end of a session (close or send AppTracker1 to the background),
\end{itemize}

\item Other states:
\begin{itemize} 
\item \TaC: terms and conditions page, 
\item \Main: main menu screen, 
\item \Settings: settings options, 
\item \Feedback: screen for giving feedback, 
\item \Info: information about the app, 
\item \Task: feedback question chosen from the \Feedback.
\end{itemize}
\end{itemize}

The new states in AppTracker 2 are the following:
\begin{itemize}
\item \OverallAll: summary statistics about the overall device usage since installing AppTracker2, 
\item \ExploreData: three options for more in-depth exploration of all recorded data,
\item \AppsToday: summaries of the current day's usage of apps,
\item \AppsbyPeriod: usage statistics for various apps by selected time period,
\item \ChartAppsInPeriod: detailed app usage when selected from \AppsbyPeriod,
\item \OverallbyApp: usage summary for a selected app,
\item \ChartOverallbyApp: detailed app usage, when selected from \OverallbyApp.
\end{itemize}

%% file: pctl_background.tex
\section{Probabilistic Temporal Logic}\label{pctl_background} 

Probabilistic Computation Tree Logic (PCTL) and its extension PCTL*
are logics that allow expression of  a probability measure of the satisfaction of a temporal
property by a state of a discrete-time Markov model.
Their syntax is the following:
\begin{center}
  \begin{tabular}{rl} 
    {\em State formulae} & $\quad \Phi ::=
    \mathit{true} \mid a \mid \lnot\, \Phi \mid \Phi \land \Phi \mid
    \Probs_{\bowtie\, p}[\Psi] \mid \Probss_{\bowtie\, p}[\Phi]$
    \\
    {\em PCTL path formulae} & $\quad \Psi ::= \TX\, \Phi \mid \Phi\,
    \TU^{\leq N}\, \Phi$
    \\
    {\em PCTL* path formulae} & $\quad \Psi ::= \Phi\ \mid \Psi \land
    \Psi \mid \lnot\, \Psi \mid \TX\, \Psi \mid \Psi\, \TU^{\leq N}\,
    \Psi$
\end{tabular}
\end{center}
where $a$ represents an atomic proposition, $\bowtie\,\in \{\leq, <,
\geq, > \}$, 
$p\in [0,1]$, and 
$N\in \mathbb{N}\cup 
\{\infty\}$.  

PCTL and PCTL* formulae (or properties) are interpreted over states of
a DTMC, with state formulae $\Phi$ evaluated over states and path
formulae $\Psi$ over paths.  We say that a DTMC satisfies a state
formulae $\Phi$ if the initial state of model satisfies $\Phi$. We
denote by $s\models \Phi$ that state $s$ satisfies $\Phi$ (or $\Phi$
is evaluated to true in state $s$). Then $s\models \mathit{true}$ is
always true; $s\models a$ iff $a$ is an atomic proposition labelling
$s$; $s\models \lnot \Phi$ iff $s\models \Phi$ is false; $s\models
\Phi_1\land\Phi_2$ iff $s\models \Phi_1$ and $s\models \Phi_2$;
$s\models\Probs_{\bowtie\, p}[\Psi]$ iff the probability that $\Psi$
is satisfied by the paths starting from state $s$ meets the bound
$\bowtie p$; $s\models \Probss_{\bowtie\, p}[\Phi]$ iff the
steady-state (long-run) probability of being in a state that satisfies
$\Psi$ meets the bound $\bowtie p$. The operators $\TX$ and $\TU$ are
called the ne{\bf X}t and the {\bf U}ntil operators respectively.
Informally, the path formulae $\TX\, \Phi$ is true on a path starting
in $s$ iff $\Phi$ is satisfied in the {\em next} state following $s$
in the path, whereas $\Phi_1\, \TU^{\leq N}\, \Phi_2$ is true on a
path $\omega$ iff $\Phi_2$ is satisfied within $N$ time-steps and
$\Phi_1$ is true up {\em until} that point.

The syntax above includes only a minimal set of operators; the
propositional operators $\mathit{false}$, disjunction $\lor$ and
implication $\implies$ can be derived. Two common derived path
operators are: the \emph{eventually} operator $\TF$ where $\TF^{\leq
  n}\,\Phi\equiv \mathit{true}\, \TU^{\leq n}\,\Phi$ and the
\emph{always} operator $\TG$ where $\TG\,\Psi \equiv
\lnot(\TF\,\lnot\,\Psi)$.  If $N =\infty$, i.e., the until operator
$\TU$ is not bounded, then the superscript is omitted.

%% file: appendix.tex
\section{Analysis Results for GPAM(3) Models of AppTracker1 and AppTracker2}\label{appendix_gpam3}

\input{appendix_AT1_GPAM3}

\input{appendix_AT2_GPAM3}

\input{appendix_GPAM3_StateToPattern}

%% file: appendix_AT1_GPAM3.tex

Similar to the GPAM(2) inductive coding process, we instantiate the
probabilistic temporal properties \linebreak \VisitProbInit, \VisitCountInit, and \StepCountInit\
from Table~\ref{tbl:properties} with all states for the
GPAM(3) models. The results of model checking these properties
instantiated with the states \OverallUsage, \LastSevenDays,
\SelectPeriod, \Stats, and \AppsInPeriod, as well as the session
counts and lengths, are given in 
Table~\ref{table:AT1_GPAM3_SAP_InitProps_4statesL1}.

Table~\ref{table:GPAM3_initial_AP_categ_AT1} shows our workings for
the inductive coding process, and Table~\ref{table:AT1_APlabels_GPAM3}
the activity pattern labels. Again, there is a strong correlation
between session length and frequency i.e. few/long and
many/short. However, we uncovered another state group that we call
\TopLevelMenu, consisting of states reachable within one-two button
taps from the main menu. Overall, session categories are finer-grained
for GPAM(3) than for GPAM(2). We note again differences between early
days usage and experienced usage, though the distinctions are not as
stark as in GPAM(2). In the second and third months we see more
\Glancing\ and \Focussing\ behaviours, and no \Browsing.  In the
second month of usage of GPAM(3) we note that two patterns have the
same label, \Glancing\Specific: they both
abstract a similar \Glancing\ behaviour, however centred on slightly
different lists of predominant states, $\{4,6,5\}$ and $\{4,5\}$.

\begin{table}[!t]
  \center 
\caption{AppTracker1. GPAM(3). Analysis of properties \VisitProbInit,
    \VisitCountInit, \StepCountInit\ for \OverallUsage, \LastSevenDays, \SelectPeriod, \Stats, and \AppsInPeriod, followed by  characteristics of session (properties \VisitProbInit\ for \UseStop, \SessionCount\ and \SessionLength) for $N=50$.
    \textcolor{blue}{blue} indicates best result,
    \textcolor{purple}{purple} indicates  middle result,
    \textcolor{orange}{orange} indicates worst result}
    \label{table:AT1_GPAM3_SAP_InitProps_4statesL1} 
    {
\begin{tabular}{c}      
\begin{tabular}{|c|c|rrr|rrr|rrr|rrr|rrr|}
  \hline 
  {\bf \scriptsize Prop.} & {\bf \scriptsize Time} &
  \multicolumn{3}{c|}{\bf \scriptsize \OverallUsage} &
  \multicolumn{3}{c|}{\bf \scriptsize \LastSevenDays} &
  \multicolumn{3}{c|}{\bf \scriptsize \SelectPeriod} &
  \multicolumn{3}{c|}{\bf \scriptsize \Stats} & \multicolumn{3}{c|}{\bf
    \scriptsize \AppsInPeriod} \\  
  \cline{3-5} \cline{6-8} \cline{9-11}
  \cline{12-14} \cline{15-17} & {\bf \scriptsize int.} & 
  {\bf \scriptsize AP1} & {\bf \scriptsize AP2} & {\bf \scriptsize AP3}
  & {\bf \scriptsize AP1} & {\bf \scriptsize AP2} & {\bf \scriptsize
    AP3} & {\bf \scriptsize AP1} & {\bf \scriptsize AP2} & {\bf
    \scriptsize AP3} & {\bf \scriptsize AP1} & {\bf \scriptsize AP2} &
  {\bf \scriptsize AP3} & {\bf \scriptsize AP1} & {\bf \scriptsize AP2}
  & {\bf \scriptsize AP3} \\ 
  \hline\hline
  \multirow{5}{*}{\begin{sideways}{\scriptsize\VisitProbInit}\end{sideways}} 
  & [0,1] & 
  \textcolor{blue}{0.99}  & \textcolor{orange}{0.78} & \textcolor{blue}{0.99} 
  & \textcolor{orange}{0.52} & \textcolor{purple}{0.91} & \textcolor{blue}{0.98} 
  & \textcolor{orange}{0.01} & \textcolor{blue}{0.90} & \textcolor{purple}{0.29} 
  & \textcolor{blue}{0.99} & \textcolor{purple}{0.85} & \textcolor{orange}{0.66} 
  & \textcolor{blue}{0.03} & \textcolor{orange}{0.00} & \textcolor{purple}{0.01} \\ 
& [0,7] & \textcolor{blue}{0.99} & \textcolor{orange}{0.62} & \textcolor{purple}{0.98} & \textcolor{orange}{0.27} & \textcolor{purple}{0.69} & \textcolor{blue}{0.98} & \textcolor{orange}{0.08} & \textcolor{purple}{0.79} & \textcolor{blue}{0.94} & \textcolor{purple}{0.77} & \textcolor{orange}{0.45} & \textcolor{blue}{0.82}  & \textcolor{orange}{0.02} & \textcolor{blue}{0.62} & \textcolor{purple}{0.53} \\  
& [0,30] 
& \textcolor{blue}{0.99} & \textcolor{orange}{0.59} & \textcolor{blue}{0.99} 
& \textcolor{orange}{0.23} &\textcolor{purple}{0.60} & \textcolor{blue}{0.99} 
& \textcolor{orange}{0.07} & \textcolor{purple}{0.65} & \textcolor{blue}{0.98} 
& \textcolor{orange}{0.16} & \textcolor{purple}{0.36} & \textcolor{blue}{0.80} 
& \textcolor{orange}{0.04} & \textcolor{purple}{0.53} & \textcolor{blue}{0.71} \\ 
& [30,60] & \textcolor{blue}{0.99} & \textcolor{orange}{0.44} & \textcolor{purple}{0.65} & \textcolor{orange}{0.78} & \textcolor{blue}{0.99} & \textcolor{purple}{0.98} & \textcolor{orange}{0.19} & \textcolor{purple}{0.71} & \textcolor{blue}{0.97} & \textcolor{blue}{0.94} & \textcolor{orange}{0.00} & \textcolor{purple}{0.25} & \textcolor{purple}{0.11} & \textcolor{blue}{0.57} & \textcolor{orange}{0.00} \\
& [60,90] & \textcolor{blue}{0.99} & \textcolor{purple}{0.90} & \textcolor{orange}{0.76} & \textcolor{orange}{0.40} & \textcolor{blue}{0.99} & \textcolor{purple}{0.74} & \textcolor{orange}{0.11} & \textcolor{purple}{0.37} & \textcolor{blue}{0.99} & \textcolor{blue}{0.95} & \textcolor{purple}{0.55} & \textcolor{orange}{0.40} & \textcolor{orange}{0.09} & \textcolor{purple}{0.33} & \textcolor{blue}{0.88} \\ 
\hline
\multirow{5}{*}{\begin{sideways}\scriptsize\VisitCountInit\end{sideways}}  
& [0,1] & \textcolor{blue}{18.89} & \textcolor{purple}{5.44} & \textcolor{orange}{1.80} & \textcolor{orange}{0.73} & \textcolor{blue}{3.76} & \textcolor{purple}{2.31} & \textcolor{orange}{0.01} & \textcolor{purple}{0.34} & \textcolor{blue}{2.18} & \textcolor{blue}{7.39} & \textcolor{orange}{1.79} & \textcolor{purple}{1.84} & \textcolor{blue}{0.33} & \textcolor{purple}{0.01} & \textcolor{orange}{0.00} \\ 
& [0,7] & \textcolor{blue}{20.47} & \textcolor{orange}{1.20} & \textcolor{purple}{6.05} & \textcolor{orange}{0.33} & \textcolor{purple}{1.13} & \textcolor{blue}{5.24} & \textcolor{orange}{0.12} & \textcolor{purple}{2.11} & \textcolor{blue}{2.72} & \textcolor{purple}{1.96} & \textcolor{orange}{0.60} & \textcolor{blue}{2.15} & \textcolor{orange}{0.04} & \textcolor{purple}{1.99} & \textcolor{blue}{5.04}\\ 
& [0,30] 
& \textcolor{blue}{22.66} & \textcolor{orange}{1.12} & \textcolor{purple}{5.37}  
& \textcolor{orange}{0.27} & \textcolor{purple}{0.89} & \textcolor{blue}{5.94} 
& \textcolor{orange}{0.10} & \textcolor{purple}{1.80} & \textcolor{blue}{3.48} 
& \textcolor{orange}{0.26} & \textcolor{purple}{0.45} & \textcolor{blue}{1.92} 
& \textcolor{orange}{0.06} & \textcolor{purple}{2.88} & \textcolor{blue}{4.07} \\ 
& [30,60] & \textcolor{blue}{16.75} & \textcolor{orange}{0.76} & \textcolor{purple}{1.12} & \textcolor{orange}{1.57} & \textcolor{blue}{9.38} & \textcolor{purple}{3.76} & \textcolor{orange}{0.23} & \textcolor{purple}{1.35} & \textcolor{blue}{3.49} & \textcolor{blue}{4.11} & \textcolor{orange}{0.00} & \textcolor{purple}{0.35} & \textcolor{purple}{1.31} & \textcolor{blue}{6.85} & \textcolor{orange}{0.00} \\
& [60,90] & \textcolor{blue}{12.84} & \textcolor{orange}{2.79} & \textcolor{purple}{4.00} & \textcolor{orange}{0.50} & \textcolor{blue}{8.16} & \textcolor{purple}{1.58} & \textcolor{purple}{0.51} & \textcolor{orange}{0.44} & \textcolor{blue}{5.59} & \textcolor{blue}{4.10} & \textcolor{orange}{0.89} & \textcolor{purple}{1.25} & \textcolor{orange}{0.43} & \textcolor{purple}{3.85} & \textcolor{blue}{7.57} \\ 
\hline\multirow{5}{*}{\begin{sideways}\scriptsize\StepCountInit\end{sideways}} 
& [0,1] & \textcolor{blue}{4.72} & \textcolor{purple}{6.04} & \textcolor{orange}{32.91} & \textcolor{orange}{68.57} & \textcolor{blue}{13.63} & \textcolor{purple}{20.50} & \textcolor{orange}{3350.93} & \textcolor{purple}{148.31} & \textcolor{blue}{21.22} & \textcolor{blue}{9.70} & \textcolor{orange}{46.07} & \textcolor{purple}{26.09} & \textcolor{blue}{1415.69} & \textcolor{grey}{---} & \textcolor{grey}{---} \\ 
& [0,7] & \textcolor{grey}{---} & \textcolor{orange}{51.43} & \textcolor{purple}{7.45} & \textcolor{grey}{---} & \textcolor{orange}{42.73} & \textcolor{purple}{10.20} & \textcolor{grey}{---} & \textcolor{orange}{32.43} & \textcolor{purple}{18.05} & \textcolor{grey}{---} & \textcolor{orange}{82.57} & \textcolor{purple}{27.66} & \textcolor{grey}{---} & \textcolor{blue}{51.56} & \textcolor{purple}{64.83}\\ 
& [0,30] 
& \textcolor{blue}{5.01} & \textcolor{orange}{55.98} & \textcolor{purple}{8.59} 
& \textcolor{orange}{247.66} & \textcolor{purple}{55.63} & \textcolor{blue}{8.68} 
& \textcolor{orange}{1153.58} & \textcolor{purple}{47.65} & \textcolor{blue}{13.44} 
& \textcolor{orange}{295.84} & \textcolor{purple}{111.09} & \textcolor{blue}{29.52} 
& \textcolor{orange}{2670.48} & \textcolor{purple}{65.95} & \textcolor{blue}{40.51} \\ 
& [30,60] & \textcolor{blue}{4.05} & \textcolor{orange}{88.28} & \textcolor{purple}{47.26} & \textcolor{orange}{44.94} & \textcolor{blue}{4.09} & \textcolor{purple}{12.66} & \textcolor{orange}{223.72} & \textcolor{purple}{39.81} & \textcolor{blue}{14.48} & \textcolor{purple}{21.93} & \textcolor{grey}{---} & \textcolor{orange}{172.74} & \textcolor{purple}{403.76} & \textcolor{blue}{59.48} & \textcolor{grey}{---} \\ 
& [60,90] & \textcolor{blue}{2.27} & \textcolor{purple}{21.54} & \textcolor{orange}{34.66} & \textcolor{orange}{98.96} & \textcolor{blue}{5.05} & \textcolor{purple}{36.49} & \textcolor{orange}{416.06} & \textcolor{purple}{106.05} & \textcolor{blue}{8.39} & \textcolor{blue}{16.21} & \textcolor{purple}{63.78} & \textcolor{orange}{96.67} & \textcolor{orange}{489.26} & \textcolor{purple}{123.08} & \textcolor{blue}{22.93} \\  
\hline
\end{tabular}
\\
\\
\\
\begin{tabular}{|c|rrr|rrr|rrr|}
  \hline
  {\bf \scriptsize Time} & \multicolumn{3}{c|}{\VisitProbInit} & \multicolumn{3}{c|}{\SessionCount} & \multicolumn{3}{c|}{\SessionLength} \\   
  \cline{2-4} \cline{5-7} \cline{8-10} {\bf \scriptsize interval} & {\bf \scriptsize AP1} & {\bf \scriptsize AP2} &
  {\bf \scriptsize AP3} & {\bf \scriptsize AP1} & {\bf \scriptsize AP2} &
  {\bf \scriptsize AP3} & {\bf \scriptsize AP1} & {\bf \scriptsize AP2} &
  {\bf \scriptsize AP3} \\  
\hline\hline
  $[0,1]$ & 0.16 & 0.99 & 0.02  & 0.18 & 12.36 & 0.02 & 376.93 & 2.95 & 2469.47  \\ 
  $[0,7]$ & 0.27 & 0.99 & 0.89 & 0.32 & 11.91 & 1.82 & --- & 3.12 & 24.40\\ 
  $[0,30]$ & 0.30 & 0.99 & 0.97 & 0.34 & 11.74 & 2.54 & 146.32 & 3.25 & 17.16 \\ 
  $[30,60]$ & 0.70 & 0.99 & 0.99 & 1.22 & 8.83 & 8.58 & 57.66 & 4.71 & 4.68 \\ 
  $[60,90]$ & 0.99 & 0.99 & 0.96 & 8.56 & 8.75 & 3.57 & 4.65 & 4.65 & 13.47  \\ 
  \hline
\end{tabular}
\end{tabular}
}
\end{table}

\begin{table}[!t]
  \center
  \caption{AppTracker1. GPAM(3). Inductive coding for labelling the
    activity patterns: categorisation of the average session counts
    and lengths, subsequently used for the initial categorisation of
    activity patterns alongside the predominant 
    states, and description of the usage intensities.  Session length 
    "--" indicates   the reachability probability for the end of
    session is less than 1 (hence infinite length). State identifiers
    are:  
    3 is \OverallUsage, 4 is \LastSevenDays, 5 is \SelectPeriod, 
    6 is \AppsInPeriod, 9 is \Stats, 10 is \UBCOverallUsage, 
    12 is \UBCStats, and 14 is \UBCAppsInPeriod.  
    State identifiers are highlighted as \colorbox{lgray}{\Summary\
      states} and \colorbox{dgray}{\Specific\ states}.  
  }  
    \label{table:GPAM3_initial_AP_categ_AT1}
{\small
\begin{tabular}{c}
\begin{tabular}{|c|c|c|c|}
\hline
\multicolumn{4}{|c|}{\bf Session count categories}   \\
\cline{1-4} {\bf v.few} & {\bf few} & {\bf mid} & {\bf many}   \\
\hline
$\{0.02,0.18,0.32,0.34,1.22\}$ & $\{1.82,2.54,3.57\}$ & $\{8.56,8.58,8.75,8.83\}$ & $\{11.74,12.36\}$ \\
\hline
\end{tabular}
\\
\\
\begin{tabular}{|c|c|c|c|c|c|}
\hline
\multicolumn{6}{|c|}{\bf Session length categories}  \\
\cline{1-6} {\bf v.short} & {\bf short} & {\bf mid} & {\bf long} & {\bf v.long} & (n/a)  \\
\hline
$\{2.95,3.12,3.25\}$ & $\{4.65,4.68,4.71\}$ & $\{13.47\}$ & $\{17.16,24.40,57.66\}$ & $\{146.32, 376.93,2469.47\}$ & -- \\
\hline
\end{tabular}
\\
\\
\begin{tabular}{|c|cc|cc|cc|}
\hline
{\bf Time} & \multicolumn{2}{c|}{\bf AP1} & \multicolumn{2}{c|}{\bf AP2} & \multicolumn{2}{c|}{\bf AP3} \\
\cline{2-3} \cline{4-5} \cline{6-7} {\bf int.} & {\bf Sessions} & {\bf State ids}
& {\bf Sessions} & {\bf State ids} & {\bf Sessions} & {\bf State ids} \\ 
\hline\hline
$[0,1]$ & v.few/v.long & \colorbox{lgray}{3},\colorbox{lgray}{9},\colorbox{lgray}{10},\colorbox{lgray}{12} & many/v.short & \colorbox{dgray}{4},\colorbox{dgray}{5},\colorbox{lgray}{9} & v.few/v.long & \colorbox{lgray}{3},\colorbox{dgray}{4},\colorbox{lgray}{9}  \\
\hline
$[0,7]$ & v.few/-- & \colorbox{lgray}{3},\colorbox{lgray}{10} & many/v.short & \colorbox{dgray}{5},\colorbox{dgray}{6},\colorbox{lgray}{3},\colorbox{dgray}{4} & few/long &  \colorbox{lgray}{3},\colorbox{dgray}{4},\colorbox{dgray}{5},\colorbox{lgray}{9},\colorbox{dgray}{6} \\
\hline
 $[0,30]$ & v.few/v.long & \colorbox{lgray}{3},\colorbox{lgray}{10} & many/v.short & \colorbox{dgray}{6},\colorbox{dgray}{5},\colorbox{dgray}{14} & few/long &  \colorbox{dgray}{4},\colorbox{lgray}{3},\colorbox{dgray}{5},\colorbox{lgray}{9},\colorbox{dgray}{6} \\
\hline
 $[30,60]$ & few/long & \colorbox{lgray}{3},\colorbox{lgray}{9},\colorbox{lgray}{12} & mid/short & \colorbox{dgray}{4},\colorbox{dgray}{6},\colorbox{dgray}{5} & mid/short & \colorbox{dgray}{4},\colorbox{dgray}{5} \\
\hline
 $[60,90]$ & mid/short & \colorbox{lgray}{3},\colorbox{lgray}{9},\colorbox{lgray}{10},\colorbox{lgray}{12} & mid/short & \colorbox{dgray}{4},\colorbox{dgray}{6} & few/mid & \colorbox{dgray}{6},\colorbox{dgray}{14} \\
\hline
\end{tabular}
\\
\\
\begin{tabularx}{\linewidth}{rcX}
  \toprule
  {\bf Usage intensity} & & {\bf Usage intensity characteristics}\\
  \midrule \Browsing & & very few/very long or few/long sessions
  centred on at least three states from the \TopLevelMenu\
  (i.e.~reachable within one-two button taps from the main menu),
  sometimes from different state groups \\ 
  \Glancing & & many/very short or mid/short sessions centred on one or
  two states \\ 
  \Focussing & & mid/mid, few/mid, few/long, very few/--, very few/very
  long sessions centred on states from the same group \\ 
  \bottomrule
\end{tabularx}
\end{tabular}
}
\end{table} 

\begin{table}[!ht]
\centering
\caption{AppTracker1. Activity pattern labels in GPAM(3) models. 
}
\label{table:AT1_APlabels_GPAM3}
{\small
  \begin{tabular}{clll}
    \toprule
    {\bf Time int.}  & {\bf AP1} & {\bf AP2} & {\bf AP3}  \\
    \midrule
    $[0,1]$ & \Focussing\Summary & \Glancing\Specific & \Browsing\Summary  \\ 
    $[0,7]$ & \Focussing\Summary & \Glancing\Specific & \Browsing\TopLevelMenu \\
    $[0,30]$ & \Focussing\Summary & \Glancing\Specific & \Browsing\TopLevelMenu \\
    $[30,60]$ & \Focussing\Summary & \Glancing\Specific & \Glancing\Specific \\
    $[60,90]$ & \Glancing\Summary & \Glancing\Specific & \Focussing\Specific \\
    \bottomrule
  \end{tabular}
}
\end{table}

%% file: appendix_AT2_GPAM3.tex

The analysis results of GPAM(3) models for AppTracker2 are listed in Table~\ref{table:GPAM3_initial_AP_categ_AT2}:
the categories for the session count and the session length, the
initial categorisation of the patterns based on session
characteristics and predominant states and the usage intensities,
which are similar to those for AppTracker1.  As before "--" indicates
the probability to reach the state \UseStop\ is less than 1, hence the
cumulated reward (step count) is infinity.  The GPAM(3) activity
pattern labels are given in Table~\ref{table:AT2_APlabels_GPAM3}.
Because \Glancing\ behaviours occur more often than in AppTracker1,
and to highlight the difference between centred on one or two states,
the notation \Focussing(\OverallAll) indicates a pattern
\Focussing\Summary\ that is centred only on the state \OverallAll.
In summary, in GPAM(3) models of AppTracker2 we uncovered the following six distinct activity patterns (compared with four in AppTracker1): \Focussing\Summary,  \Focussing\SpecificByPeriod, \Glancing\Summary, \Focussing(\OverallAll), \Glancing(\Main), 
and \Browsing\Summary.

\begin{table}[!ht] 
  \center
  \caption{AppTracker2. GPAM(3). Categorisation of the average session counts and lengths, subsequently used for the 
  initial categorisation of activity patterns alongside the predominant states, and description of the usage intensity labels. 
  State identifiers (ids) are: 1 is \Main, 5 is \OverallAll, 6 is \ExploreData, 
    8 is \AppsToday, 9 is \SelectPeriod, 10 is \Stats, 
    11 is \AppsbyPeriod, 12 is \UBCAppsInPeriod, 13 is \OverallbyApp, 
    14 is \UBCStats, 15 is \UBCOverallbyApp. State identifiers are highlighted as \colorbox{lgray}{\Summary\ states} and \colorbox{dgray}{\Specific\ states}.}  
    \label{table:GPAM3_initial_AP_categ_AT2} 
{\small
\begin{tabular}{c}
\begin{tabular}{|c|c|c|c|}
\hline
\multicolumn{4}{|c|}{\bf Session count categories}   \\
\cline{1-4} {\bf v.few} & {\bf few} & {\bf mid} & {\bf many}   \\
\hline
$\{0.06,0.09\}$ & $\{1.62,2.24,2.77,2.95,3.56\}$ & $\{5.56,6.69,6.90,7.47,8.21,8.78,9.20\}$ & $\{14.80\}$ \\
\hline
\end{tabular}
\\
\\
\begin{tabular}{|c|c|c|c|c|}
\hline
\multicolumn{5}{|c|}{\bf Session length categories}  \\
\cline{1-5}{\bf v.short} & {\bf short} & {\bf mid} & {\bf long} & (n/a) \\
\hline
$\{2.32\}$ & $\{4.25,4.97\}$ & $\{5.62,6.49,6.53,7.02,7.86,12.29\}$ & $\{18.32,23.14,31.37\}$ & -- \\
\hline
\end{tabular}
\\
\\
\begin{tabular}{|c|cc|cc|cc|}
\hline
{\bf Time} & \multicolumn{2}{c|}{\bf AP1} & \multicolumn{2}{c|}{\bf AP2} & \multicolumn{2}{c|}{\bf AP3} \\
\cline{2-3} \cline{4-5} \cline{6-7} {\bf int.} & {\bf Sessions} & {\bf State ids}
& {\bf Sessions} & {\bf States ids} & {\bf Sessions} & {\bf State ids} \\ 
\hline\hline
$[0,1]$ & mid/mid & \colorbox{dgray}{6},\colorbox{dgray}{10},\colorbox{dgray}{13},\colorbox{dgray}{15} & v.few/-- & \colorbox{lgray}{5} & few/long & \colorbox{lgray}{5},\colorbox{lgray}{8},\colorbox{dgray}{6}  \\
\hline
$[0,7]$ & few/long & \colorbox{lgray}{5},\colorbox{lgray}{8},\colorbox{dgray}{6} & v.few/-- & \colorbox{lgray}{5} & mid/mid & \colorbox{dgray}{6},\colorbox{dgray}{13},\colorbox{dgray}{10},\colorbox{dgray}{15}  \\
\hline
$[0,30]$ & mid/short & \colorbox{lgray}{5},\colorbox{lgray}{8} & few/long & \colorbox{lgray}{5},\colorbox{lgray}{8},\colorbox{dgray}{6} & mid/mid & \colorbox{dgray}{9},\colorbox{dgray}{6},\colorbox{dgray}{11}  \\
\hline
$[30,60]$ & mid/mid & \colorbox{lgray}{5},\colorbox{lgray}{8} & mid/short & \colorbox{lgray}{5},\colorbox{lgray}{8} & few/mid & \colorbox{dgray}{11},\colorbox{dgray}{9},\colorbox{dgray}{6},\colorbox{dgray}{13}  \\
\hline
$[60,90]$ & few/-- & \colorbox{lgray}{5},\colorbox{lgray}{8}
& mid/mid & \colorbox{lgray}{5},\colorbox{lgray}{8} & many/v.short & 1 \\
\hline
\end{tabular}
\\
\\
\begin{tabularx}{0.9\linewidth}{rX}
\toprule
{\bf Usage intensity} & {\bf Usage intensity characteristics}\\
\midrule
\Browsing & few/long sessions centred on at least three states reachable within 
one-two button taps from the main menu (such as \Summary\ and \SpecificMenu\ states), sometimes from different state groups \\ 
\Glancing & mid/short or many/very short sessions centred on one or two states \\
\Focussing & mid/mid, few/mid, few/--, or very few/-- sessions centred on states from the same group. \\
\bottomrule
\end{tabularx}
\end{tabular}
}
\end{table}

\begin{table}[!t]
\centering
\caption{AppTracker2. Activity pattern labels in GPAM(3) models.}
\label{table:AT2_APlabels_GPAM3}
{\small
  \begin{tabular}{clll}
    \toprule
    {\bf Time int.}  & {\bf AP1} & {\bf AP2} & {\bf AP3}  \\
    \midrule
    $[0,1]$ & \Focussing\SpecificByApp & \Focussing(\OverallAll) & \Browsing\MyTopApps \\ 
    $[0,7]$ & \Browsing\MyTopApps & \Focussing(\OverallAll)  & \Focussing\SpecificMenu \\ 
    $[0,30]$ & \Glancing\MyTopApps & \Browsing\MyTopApps & \Focussing\SpecificByPeriod \\ 
    $[30,60]$ & \Focussing\MyTopApps & \Glancing\MyTopApps & \Focussing\SpecificByPeriod \\ 
    $[60,90]$ & \Focussing\MyTopApps & \Focussing\MyTopApps & \Glancing(\Main) \\ 
    \bottomrule
  \end{tabular}
}
\end{table}

%% file: appendix_GPAM3_StateToPattern.tex

\begin{figure*}[!t]
  \centering 
  \subfigure[AppTracker1]{\includegraphics[width=0.9\textwidth]{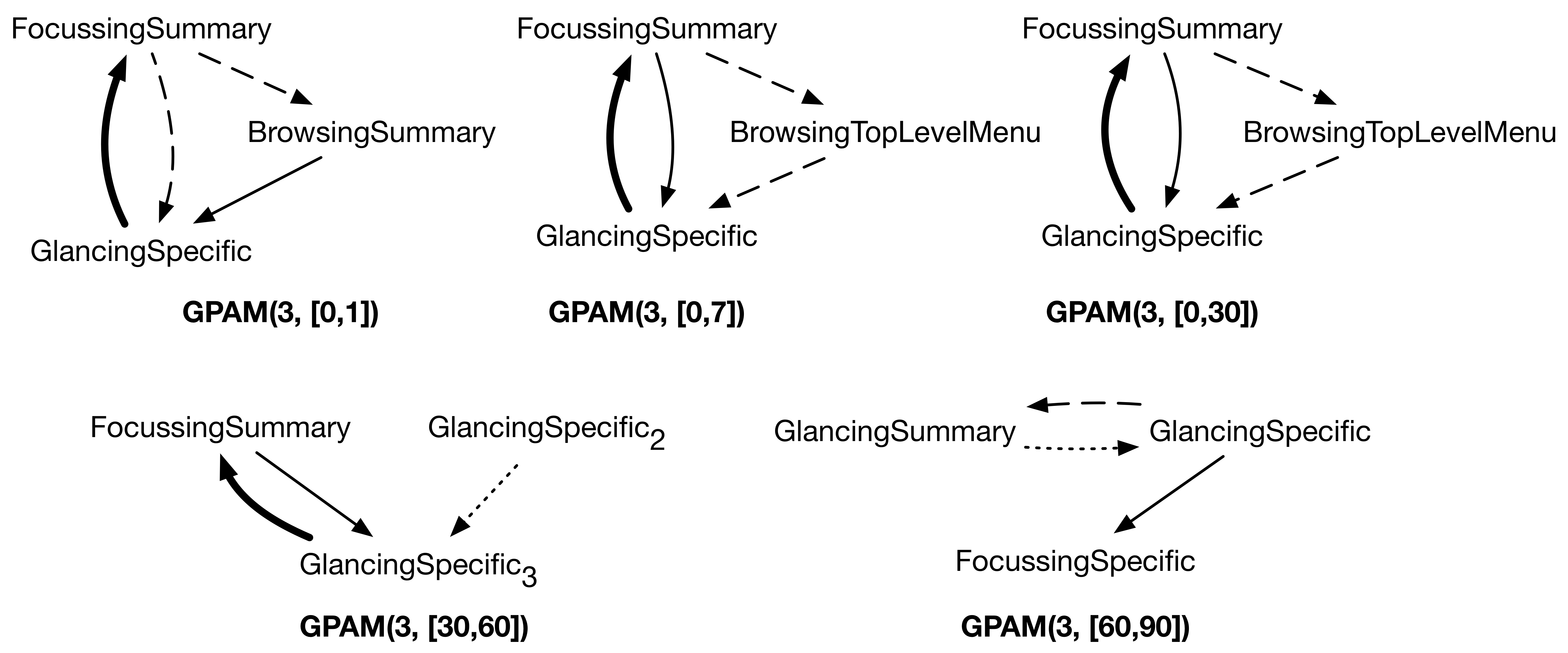}\label{fig:at1_gpam3_state2pattern}}
  \vspace{.9cm}
  
  \subfigure[AppTracker2]{\includegraphics[width=0.9\textwidth]{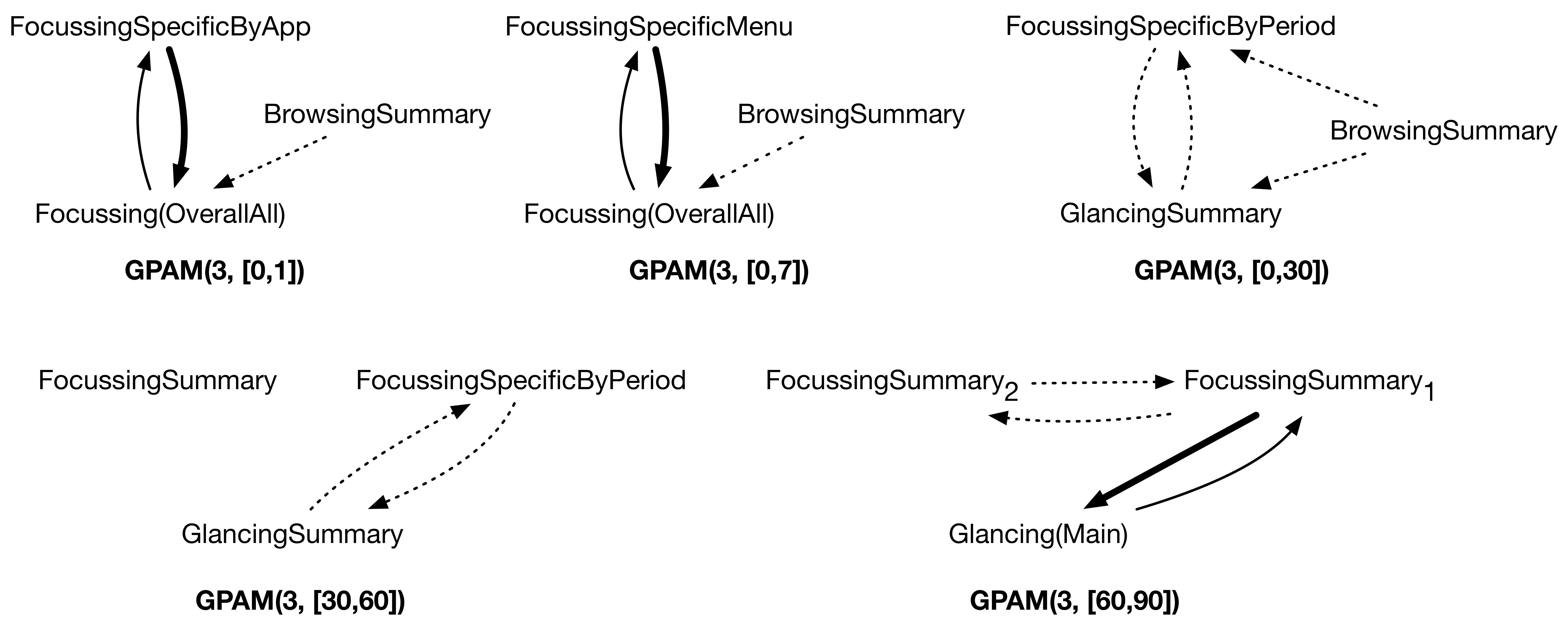}\label{fig:at2_gpam3_state2pattern}} 
  \caption{AppTracker1 and AppTracker2. Visualisations  of \StateToAPProb: the average
    likelihood of changing activity pattern from an observed state
    within a session in GPAM(3) models for different time
    intervals. Excepting the \Stats\ and \UBCStats\ states for which
    the property returns lower probabilities than the average, the
    probabilities returned for all states are very similar.  
    Dotted arrows denote probabilities in $[0.02,0.2)$, dashed arrows in $(0.2,0.5)$, thin arrows in $[0.5,0.8)$, 
    and thick arrows $\geq 0.8$. Likelihoods lower than 0.02 are omitted. 
    For GPAM(3,[30,60]) in AppTracker1 we distinguish the two identically-labelled patterns with subscript $2$ or $3$,  indicating pattern identifier (AP2 or AP3). }
\label{fig:at1_at2_gpam3_state2pattern}
\end{figure*}

For GPAM(3) models (see Fig.~\ref{fig:at1_gpam3_state2pattern}), for
early days users, \Glancing\Specific\ almost always leads to
\Focussing\Summary\
(except when in the \Stats\ state where the probability is lower; this
is not indicated in Fig.~\ref{fig:at1_gpam3_state2pattern}, which
gives averages).  In the second month, it is likely to transition
between \Glancing\Specific$_3$ (centred on \LastSevenDays\ and
\SelectPeriod) and \Focussing\Summary.  
In the third month of usage there is a very low probability of
transitioning from \Glancing\Summary\ or from \Focussing\Specific\ to
another activity pattern, compared to \Glancing\Specific.

\begin{figure*}[!t]
  \centering 
  \subfigure[AppTracker1]{\includegraphics[width=0.47\textwidth]{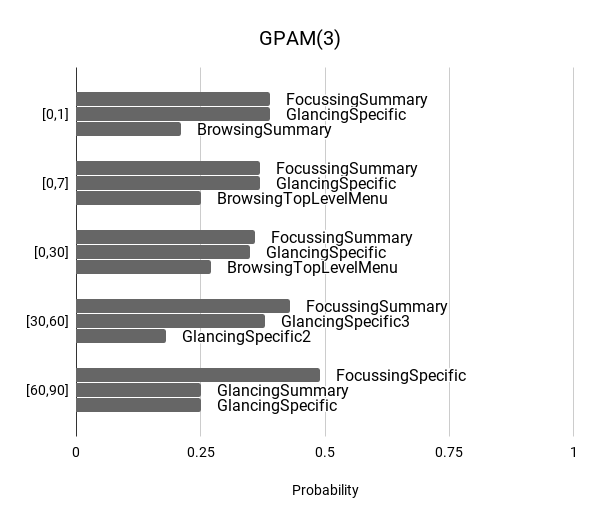}\label{fig:chart_longrun_GPAM3_AT1}}
   \quad
  \subfigure[AppTracker2]{\includegraphics[width=0.47\textwidth]{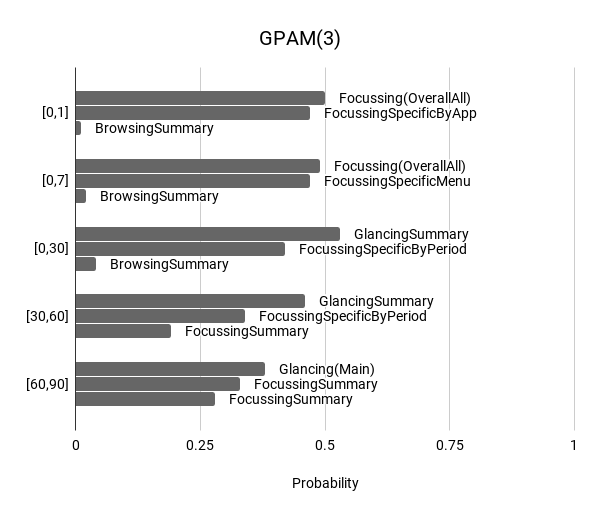}\label{fig:chart_longrun_GPAM3_AT2}
   
  } 
  \caption{AppTracker1 and AppTracker2. Visualisation of probabilities of being in each
    activity pattern in the long run in GPAM(3) models for
    different time intervals; the entries are ordered decreasingly for each time interval. 
    }
\label{fig:steadystate_AT1_AT2}
\end{figure*}

Figure~\ref{fig:steadystate_AT1_AT2} shows the results of the \LongRunPattern\ property for different GPAM(3) models for AppTracker1 and AppTracker2. 
In the early days usage models, \Focussing\Summary\ and
\Glancing\Specific\ are equally likely and prevailing, compared with
\Browsing\Summary\ and \Focussing\Specific.  In the second month, the
probability of being in \Glancing\Specific\ centred on \LastSevenDays\
and \SelectPeriod\ is higher than being in \Glancing\Specific\ centred
on \LastSevenDays, \AppsInPeriod, and \SelectPeriod\ (0.38 compared to
0.18, respectively); however, overall \Glancing\ behaviour in the
\LastSevenDays\ and \SelectPeriod\ sub-menus is more prevalent than
\Focussing\ behaviour centred on the (typically exploratory) \Summary\
states \OverallUsage\ and \Stats.  In the third month,
\Focussing\Specific\ prevails, as in GPAM(2).  However we see for the
first time \Glancing\ between \OverallUsage\ and \Stats, which could
indicate that while the sessions may be short, they may not be
exploratory but indicate more purposeful user intent.
We conclude that viewing of \Specific\ states becomes a more
\Focussing\ behaviour for experienced usage.

